\documentclass[aps,prb,twocolumn,showpacs]{revtex4}
\usepackage{amssymb}
\usepackage{graphicx}
\usepackage{amsmath}
\usepackage{color}
\usepackage{bm}

\usepackage[normalem]{ulem}

\begin{document}
\newcommand {\beq} {\begin{equation}}
\newcommand {\eeq} {\end{equation}}
\newcommand {\bqa} {\begin{eqnarray}}
\newcommand {\eqa} {\end{eqnarray}}
\newcommand {\ba} {\ensuremath{b^\dagger}}
\newcommand {\Ma} {\ensuremath{M^\dagger}}
\newcommand {\psia} {\ensuremath{\psi^\dagger}}
\newcommand {\psita} {\ensuremath{\tilde{\psi}^\dagger}}
\newcommand{\lp} {\ensuremath{{\lambda '}}}
\newcommand{\A} {\ensuremath{{\bf A}}}
\newcommand{\Q} {\ensuremath{{\bf Q}}}
\newcommand{\kk} {\ensuremath{{\bf k}}}
\newcommand{\kp} {\ensuremath{{\bf k'}}}
\newcommand{\rr} {\ensuremath{{\bf r}}}
\newcommand{\rp} {\ensuremath{{\bf r'}}}
\newcommand {\ep} {\ensuremath{\epsilon}}
\newcommand{\nbr} {\ensuremath{\langle ij \rangle}}
\newcommand {\no} {\nonumber}
\newcommand{\up} {\ensuremath{\uparrow}}
\newcommand{\dn} {\ensuremath{\downarrow}}

\newcommand{\tb}[1]{\textcolor{magenta}{#1}}


\begin{abstract}

We study the dynamics of a class of integrable non-Hermitian
free-fermionic models driven periodically using a continuous drive
protocol characterized by an amplitude $g_1$ and frequency
$\omega_D$. We derive an analytic, albeit perturbative, Floquet
Hamiltonian for describing such systems using Floquet perturbation
theory with $g_1^{-1}$ being the perturbation parameter. Our
analysis indicates the existence of special drive frequencies at
which an approximately conserved quantity emerges. The presence of
such an almost conserved quantity is reflected in the dynamics of
the fidelity, the correlation functions and the half-chain
entanglement entropy of the driven system. In addition, it also
controls the nature of the steady state of the system. We show that
one-dimensional (1D) transverse field Ising model, with an imaginary
component of the transverse field, serves as an experimentally
relevant example of this phenomenon. In this case, the transverse
magnetization is approximately conserved; this conservation leads to
complete suppression of oscillatory features in the transient
dynamics of fidelity, magnetization, and entanglement of the driven
chain at special drive frequencies. We discuss the nature of the
steady state of the Ising chain near and away from these special
frequencies, demonstrate the protocol independence of this
phenomenon by showing its existence for discrete drive protocols,
and suggest experiments which can test our theory.

\end{abstract}
\title{Emergent conservation in Floquet dynamics of integrable non-Hermitian models}

\author{Tista Banerjee and K. Sengupta }
 \affiliation{School of Physical Sciences, Indian Association for the Cultivation of
 Science, Jadavpur, Kolkata 700032, India.}

\date{\today}

\maketitle
\section{Introduction}
\label{intro}

The study of non-equilibrium dynamics of closed quantum systems has
received tremendous theoretical
\cite{rev1,rev2,rev3,rev4,rev4a,rev5,rev6,rev7,rev8} and
experimental \cite{exp1,exp2,exp3,exp4} attention in recent years.
Out of the several protocols available to drive a system out of
equilibrium, periodic drive protocols have been studied most
intensely. The evolution operator $U$ for such periodically driven
systems at stroboscopic times $t= nT$, where $T= 2 \pi/\omega_D$ is
the time period of the drive, $n$ is an integer, and $\omega_D$ is
the drive frequency, can be expressed in terms of its Floquet
Hamiltonian $H_F$ as \cite{fl1,rev8}
\begin{eqnarray}
U(nT,0) &=& \exp[-i H_F n T/\hbar] \label{evol1}
\end{eqnarray}
The study of such driven systems therefore amounts to analysis of
their Floquet Hamiltonian.

The theoretical focus on periodically driven closed quantum systems
is mostly due to the fact that they display various features that
have no analogue in their aperiodic counterparts. Some of these
include generation of topologically non-trivial Floquet states
\cite{topo1,topo2,topo3,topo4,topo5}, realization of time crystals
\cite{tc1,tc2,tc3}, and tuning ergodicity properties of
non-integrable quantum systems \cite{bm1,bm2,bm3}. In addition, they
host phenomena such as dynamical transitions
\cite{dtran1,dtran2,dtran3}, dynamical localization
\cite{dl1,dl2,dl3}, and dynamical freezing
\cite{df1,df2,df3,df4,df5,bm2,bm3}.

More recently, there has been considerable interest in study of
non-Hermitian quantum Hamiltonians\cite{nhrev,
nonhlit1,nonhlit2,nonhlit3,nonhlit4,nonhlit5,
nonhlit6,nonhlit7,nonhlit8,nonhlit9,nonhlit10,nonhlit11,nonhlit12,nonhlit13}.
Such Hamiltonians may provide effective description for open quantum
systems \cite{manas1}. In addition, they display several interesting
features such as non-Hermitian skin effect
\cite{skinherm1,skinherm2,skinherm3}, phase transition related to
explicit breaking of PT symmetry \cite{nhrev}, and the presence of
exceptional points where two complex eigenvalues of a such
Hamiltonians coincide and corresponding eigenstates coalesce
\cite{nhrev}. The presence of such exceptional points leads to
unconventional topological features and novel bulk-edge
correspondence in these systems which have no analogue in systems
described by Hermitian quantum Hamiltonians
\cite{eptop1,eptop2,eptop3,eptop4,eptop5}.

The description of out-of-equilibrium dynamics of such non-Hermitian
quantum systems has also been carried out \cite{nhdyn1,nhdyn2,
nhdyn3,nhdyn4,nhdyn5}. Most of these studies concentrated on
periodically driven systems and can be classified into two distinct
groups. The first involves study of systems driven using continuous
protocols at high frequencies where Magnus expansion may be used to
obtain analytic insight \cite{nhdyn3}. The second involves use of
discrete drive protocols where exact solution of the Floquet
Hamiltonian is available for integrable non-Hermitian models
\cite{nhdyn1,nhdyn2}. Such studies has led to several interesting
phenomena such as non-Hermitian analogue of Floquet dynamical
transitions \cite{nhdyn1}, optically induced Liftshitz transition in
non-Hermitian Weyl semimetals \cite{nhdyn3}, drive-induced PT
symmetry breaking \cite{nhdyn2}, and non-Hermitian topological
phases and transitions \cite{nhdyn1}. In addition, quench dynamics
of non-Hermitian quantum spin chains has also been studied with
focus on time evolution of correlation and entanglement entropy of
such a chain following the quench \cite{nhdyn5}.

In this work we study the periodic dynamics of a non-Hermitian
free-fermionic integrable model whose Hamiltonian is given by
\begin{eqnarray}
H &=& \sum_{\vec k} \psi_{\vec k}^{\dagger} \left(\tau_3 (g + i
\gamma -a_{3\vec k}) +\Delta_{\vec k} \tau_1 \right) \psi_{\vec k}
\label{hamint}
\end{eqnarray}
where $\psi_{\vec k}$ is a two-component fermion field and $\vec
\tau = (\tau_1,\tau_2,\tau_3)$ denotes corresponding Pauli matrices.
Here $g$, $a_{3 \vec k}$ and $\Delta_{\vec k}$ are parameters of the
model, and the presence of $\gamma>0$ makes the model non-Hermitian.
The Hermitian counterparts ($\gamma=0$) class of models serves as
prototype for a study of wide range of condensed matter system; in
$d=1$, it represents Ising and XY spin models \cite{subir1}. For
these models,  $a_{3 k}=2\cos k$, $\Delta_{k} =2\sin k$ and the two
component field $\psi_{k} = (c_{k}, c_{-k}^{\dagger})^T$ where $c_k$
denote fermion annihilation operator. In $d=2$, $H(\gamma=0)$
describes the physics of Dirac quasiparticles in graphene
\cite{graphenerev} and on surfaces of topological insulators
\cite{tirev}, as well as the fermionic description of the Kitaev
honeycomb model \cite{kitaev1}. Finally, in $d=3$, the model can be
used to describe quasiparticles in Weyl semimetals \cite{weylrev}.

The presence of a non-zero $\gamma$ leads to non-Hermitian nature of
the model. One context in which such a term naturally appears is the
1D Ising model in the presence of a measuring operator which
measures $\hat n_j = (1-\sigma_j^z)/2$ (where $\sigma_j^z$ denotes
the usual Pauli matrix representing the spin on site $j$ of the
chain) with a rate $\gamma$ and in the so-called no-click limit
\cite{dalibard1,daley1,nhdyn5}; this leads to a complex magnetic
field term in the effective Hamiltonian of the spin chain
\cite{nhdyn5}. Similar models of non-Hermitian chains have been
discussed in different contexts as well \cite{chen1,lu1}. In the
present manuscript, we shall assume the existence of such
non-hermiticity and study the Floquet dynamics of the resultant
model. We note in this context that the quench dynamics of such a
model has already been studied in Ref.\ \onlinecite{nhdyn5}.

The main results that we obtain from our study are as follows.
First, we obtain an perturbative Floquet Hamiltonian, using Floquet
perturbation theory (FPT), which reproduces all qualitative features
of the dynamics of the model and provide analytical insight into
emergence of approximate conserved quantities in this system. The
FPT uses inverse of the drive amplitude as the perturbation
parameter; it produces qualitatively accurate results both in high
and intermediate drive frequency regime  where a standard
high-frequency expansion fails \cite{rev8}.

Second, using the FPT, we identify special frequencies at which the
first order Floquet Hamiltonian of the system, $H_F^{(1)}$ leads to
conserved quantities, {\it i.e.}, $[H_F^{(1)},\hat O]=0$ for a
specific operator $\hat O$. An example of such an operator, as we
shall show, is the transverse magnetization of the Ising chain. Such
a conservation is approximate since it is violated by higher order
terms in the Floquet Hamiltonian. Nevertheless, we show, that the
approximate conservation leaves distinct imprint on the dynamics of
the system which turns out to be qualitatively different near and
away from these special frequencies. We also demonstrate the
protocol-independence of this phenomenon by demonstrating its
presence for the discrete square pulse protocol.

Third, we find that at these special frequencies, the correlation
functions, fidelity, and entanglement entropy shows distinct lack of
transient oscillations provided one starts from an eigenstate  of
the conserved operator $\hat O$. The absence of such oscillations,
which are typically present when the drive frequency is different
from the special frequencies, can be directly linked to the
approximate conservation mentioned above. Moreover, the steady state
of the driven system turns out to be close to an eigenstate of the
nearly conserved operator for any chosen initial state. For example,
consider the non-Hermitian Ising model whose Hamiltonian is given by
\begin{eqnarray}
H_{\rm Ising} &=& - J \left (\sum_{\langle i j\rangle} \sigma_i^x
\sigma_j^x + (h(t)+i \gamma) \sum_j \sigma_j^z \right) \label{isham}
\end{eqnarray}
where $J$ s the interaction strength, $\sigma^x_j$ and $\sigma^z_j$
denote Pauli matrices on site $j$, and $h(t)=h_0 + h_1 \cos \omega_D
t$ denotes the time-dependent dimensionless transverse field. The
transverse magnetization of this Ising chain is given by $S_z=\sum_j
\sigma_j^z$. At these special frequencies $S_z$ is almost conserved
and the steady state, for $\gamma>0$, is close to the ferromagnetic
state with all spins up (eigenstate of $\sigma_j^z$ with eigenvalue
$1$). Moreover the magnetization dynamics, starting from all
spin-down state, show complete absence of transient oscillations
which are normally present at other drive frequencies. Thus our
results show that the emergent approximate conservation law in such
driven system leaves its imprint on both the dynamics and the steady
state values of experimentally accessible quantities such as
magnetization of the Ising model. To the best of our knowledge, this
phenomenon has not been pointed out earlier in the literature.

The organization of the rest of the paper is as follows. In Sec.\
\ref{fpt}, we demonstrate the emergence of conserved quantities via
derivation of the Floquet Hamiltonian corresponding driven free
fermionic systems. We also provide semi-analytic expressions of
fidelity, correlation functions, and entanglement entropy for the
driven model. This is followed by Sec.\ \ref{numerics} where we
present our numerical results for the 1D transverse
field Ising model demonstrating qualitative match between results
obtained from FPT and exact numerics. Finally, in section Sec.\
\ref{diss}, we discuss our main results, suggest possible
experiments which can test our theory, and conclude. The presence of
similar emergence of approximate conserved quantities for discrete
drive protocol is discussed in the appendix.

\section{Floquet perturbation theory}
\label{fpt}

In this section, we provide an analytic, albeit perturbative
expression of the Floquet Hamiltonian of the  driven integrable
non-Hermitian model given by Eq.\ \ref{hamint} using Floquet
perturbation theory. The protocol that we use is given by
\begin{eqnarray}
g(t)= g_0 + g_1 \cos \omega_D t \label{prot1}
\end{eqnarray}
where $g_0$ is the static part of the drive and $g_1$ is the drive
amplitude. We compute the Floquet Hamiltonian in Sec.\
\ref{fptderiv}. This is followed by analytic expressions of
correlation function, fidelity, entanglement entropy for the driven
model in Sec.\ \ref{fptexp}.

\subsection{Perturbative Floquet Hamiltonian}
\label{fptderiv}

In the presence of the drive given by Eq.\ \ref{prot1}, the Floquet
Hamiltonian corresponding to Eq.\ \ref{hamint} can not be computed
exactly. This is in contrast to discrete protocols studied in the
literature \cite{nhdyn1,nhdyn2}. To obtain an analytic
understanding, we therefore use the Floquet perturbation theory to
compute $H_F$ in the regime where $g_1 \gg g_0,|\Delta_{\vec k}|,
|a_{3 \vec k}|$. In this regime, one can write the Hamiltonian as
$H_{\vec k}= H_{0 \vec k}+H_{1 \vec k}$ where
\begin{eqnarray}
H_{0 \vec k} &=& \tau_3 g_1 \cos \omega_D t \nonumber\\
H_{1 \vec k} &=& \tau_3 (g_0 + i \gamma -a_{3 \vec k}) + \tau_1
\Delta _{\vec k}.  \label{hamdiv}
\end{eqnarray}
In what follows we shall treat $H_{1 \vec k}$ perturbatively.

We begin by computing the evolution operator which, to zeroth order
in $g_1$, is given by
\begin{eqnarray}
U_{0\vec k}(t,0) &=& e^{-i \int_0^t  H_{0 \vec k} dt'/\hbar}=
\exp\left[-i \tau_3 \frac{g_1 \sin \omega_D t}{\hbar \omega_D}
\right] \label{uevol1}
\end{eqnarray}
Thus $U_{0 \vec k}(T,0) = I$ (where $I$ denotes the $2 \times 2$
identity matrix) and $H_{F \vec k}^{(0)}=0$ for all $\vec k$. Note
that the expression of $U_{0 \vec k}(t,0)$ is derived using the fact
that $H_{0 \vec k}(t)$ commutes with itself at all times.

The first order Floquet Hamiltonian can be constructed using
standard perturbation theory. To this we end, we first write the
expression of $U_{1 \vec k}(T,0)$ which is given by
\begin{eqnarray}
&& U_{1 \vec k}( T,0) = \frac{-i}{\hbar} \int_0^T dt \; U_{0 \vec
k}^{\dagger}(t,0) H_{1 \vec k} U_{0 \vec k}(t,0) \label{u1exp} \\
&& = \frac{-i T}{\hbar} \left[ \tau_3 (g_0 + i \gamma -a_{3 \vec k})
+ \tau_1 \Delta_{\vec k} J_0\left(\frac{2g_1}{\hbar \omega_D}\right)
\right], \nonumber
\end{eqnarray}
where $J_0(x)$ denotes the zeroth order Bessel function. Note that
the first term in Eq.\ \ref{u1exp} follows trivially since
$U_0(t,0)$ commutes with $\tau_3$ at all times. The computation of
the second term can be done in a straightforward manner using the
relation  $\tau_x U_{0\vec k}(t,0)= U_{0\vec
k}^{\dagger}(t,0) \tau_x$ and the identity $ \exp[i a \sin x]=
\sum_{n=-\infty}^{\infty} J_n(a) \exp[i nx]$. Using Eq.\
\ref{u1exp}, we find that the first order Floquet Hamiltonian is
given by
\begin{eqnarray}
H_{F \vec k}^{(1)} &=& \frac{i \hbar}{T}  U_{1 \vec k}(T,0) \nonumber\\
&=& \tau_3 (\alpha_{\vec k} +i \gamma) + \tau_1 \Delta_{\vec k}
J_0\left(\frac{2g_1}{\hbar \omega_D}\right) \label{fordfl}
\end{eqnarray}
where $\alpha_{\vec k}= g_0-a_{3 \vec k}$.

We note that at special frequencies, for a fixed drive amplitude,
which satisfy $2g_1/(\hbar \omega_m^{\ast})= \rho_m$ where $\rho_m$
denotes the position of the $m^{\rm th}$ zero of $J_0$, the
off-diagonal term of  $H_{F \vec k}^{(1)}$ vanishes
for all $\vec k$. At these frequencies, $[H_{F \vec k}^{(1)},
\tau_3]=0$. This constitutes an emergent dynamical symmetry which
forces the dynamics to conserve $\tau_3$ for all $\vec k$. This
symmetry will be broken by higher order terms in the Floquet
Hamiltonian as we shall show later in this section. However, we note
that at large drive frequencies, the contribution of the higher
order Floquet Hamiltonian are small and we shall see that the
correlation functions of the driven system bear signature of this
approximate dynamical symmetry. The presence of similar special
frequencies for discrete square pulse protocol has been shown in the
Appendix.

Next we compute the second order Floquet Hamiltonian. To this end,
we first note that the second order evolution operator $U_{2 \vec
k}(T,0)$ is given by
\begin{widetext}
\begin{eqnarray}
U_{2 \vec k}(T,0) &=& \left(\frac{-i}{\hbar}\right)^2 \int_0^T dt_1 \;
 U_{0 \vec k}^{\dagger}(t_1,0) H_{1 \vec k} U_{0 \vec
k}(t_1,0) \int_0^{t_1} dt_2 \;
U_{0 \vec k}^{\dagger}(t_2,0) H_{1 \vec k} U_{0 \vec k}(t_2,0) \nonumber\\
&=& \left(\frac{-i}{\hbar}\right)^2 \int_0^T dt_1 \int_0^{t_1} dt_2 \; \left( \begin{array}{cc}
(\alpha_{\vec k} + i\gamma)^2 + A_{\vec k}(t_1,t_2) & (\alpha_{\vec k} + i\gamma) B_{\vec{k}}(t_1,t_2) \\
-(\alpha_{\vec k} + i\gamma) B_{\vec k}^{\ast}(t_1,t_2) &
(\alpha_{\vec k} + i\gamma)^2 + A_{\vec k}^{\ast}(t_1,t_2)
\end{array} \right) \label{tword}
\end{eqnarray}
\end{widetext}
where the functions $A_{\vec k}(t_1,t_2)$ and $B_{\vec k}(t_1,t_2)$
are given by
\begin{eqnarray}
A_{\vec k}(t_1,t_2) &=&  \Delta^2_{\vec k} e^{\frac{2ig_1}{\hbar
\omega_D} (\sin \omega_D t_1 - \sin
\omega_D t_2)}  \label{abfn} \\
B_{\vec k}(t_1,t_2) &=& \Delta_{\vec k} \left( e^{\frac{2ig_1}{\hbar
\omega_D} \sin \omega_D t_2} - e^{\frac{2ig_1}{\hbar \omega_D} \sin
\omega_D t_1} \right) \nonumber
\end{eqnarray}
The integrations can be easily carried out using standard identities
involving Bessel functions. A straightforward computation leads to
the second order Floquet Hamiltonian
\begin{eqnarray}
H_{F \vec k}^{(2)} &=& \frac{i \hbar}{T} \left( U_{2 \vec k}(T,0)-
U_{1 \vec k}^2(T,0)/2 \right) \label{twoordfl}\\
&=& -\tau_3 4 \Delta^2_{\vec k} \sum_{n=0}^{\infty} \frac{
J_0\left(\frac{2g_1}{\hbar \omega_D}\right)
J_{2n+1}\left(\frac{2g_1}{\hbar \omega_D}\right)}{(2n+1) \hbar
\omega_D} \nonumber\\
&& + \tau_1 4 \Delta_{\vec k}(\alpha_{\vec k}+i \gamma)
\sum_{n=0}^{\infty} \frac{J_{2n+1}\left(\frac{2 g_1}{\hbar
\omega_D}\right)}{(2n+1) \hbar \omega_D} \nonumber
\end{eqnarray}
Combining Eqs.\ \ref{fordfl} and \ref{twoordfl}, we find the final
Floquet Hamiltonian to be
\begin{eqnarray}
H_{F \vec k} &=& \tau_3 S_{1 \vec k} + \tau_1  S_{2 \vec k}, \quad
 S_{1 \vec k} = (\alpha_{1 \vec k} + i \gamma) \nonumber\\
\alpha_{1 \vec k} &=& \alpha_{\vec k} -2 \Delta^2_{\vec k}
\sum_{n=0}^{\infty} \frac{ J_0\left(\frac{2g_1}{\hbar
\omega_D}\right) J_{2n+1}\left(\frac{2g_1}{\hbar
\omega_D}\right)}{(n+1/2) \hbar \omega_D} \nonumber\\
 S_{2 \vec k} &=& \Delta_{\vec k} (\alpha_{2 \vec k} + i
\gamma \lambda )
\nonumber\\
\alpha_{2 \vec k} &=& \left( J_0\left(\frac{2g_1}{\hbar
\omega_D}\right) +  \alpha_{\vec k} \lambda
\right) \nonumber\\
\lambda  &=& 2 \sum_{n=0}^{\infty} \frac{J_{2n+1}\left(\frac{2
g_1}{\hbar \omega_D}\right)}{(n+1/2) \hbar \omega_D} \label{flham}
\end{eqnarray}

The energy spectrum of the Floquet Hamiltonian can be easily found
by diagonalizing $H_{F \vec{k}}$. We find two energy bands whose
expressions are given by
\begin{eqnarray}
E_{\vec k}^{\pm} &=& \pm E_{\vec k} ;\quad
E_{\vec k}= \left(\epsilon_{\vec k} + i \Gamma_{\vec k}
\right) \label{fldisp}\\
\epsilon_{\vec k} &=& \frac{1}{\sqrt{2}} \sqrt{\beta_1+
\sqrt{\beta_1^2+ 4 \gamma^2 (\alpha_1 + \Delta_{\vec k} \alpha_2
\lambda)^2}} \nonumber\\
\Gamma_{\vec k} &=& \frac{\gamma (\alpha_1 + \Delta^2_{\vec k}
\alpha_2
\lambda)}{\epsilon_{\vec k}} \nonumber\\
\beta_1 &=&  \alpha_1^2+\Delta^2_{\vec k} \alpha_2^2-
\gamma^2\left(1+\Delta^2_{\vec k}\lambda^2\right) \nonumber
\end{eqnarray}
where we have not written down the $\vec k$ dependence of
$\alpha_1$, $\beta_1$, and $\alpha_2$ defined in Eq.\ \ref{flham}
for brevity.

We note that the Floquet quasienergy spectrum allows for long-lived
quasienergy excitations for $\vec k= \vec k_0$ which satisfies
$\alpha_{1 \vec k_0}= -\lambda \Delta^2_{\vec k_0} \alpha_{2 \vec k_0}$.
Furthermore it also shows the presence of exceptional point for a
critical $\gamma=\gamma_E$ such that
\begin{eqnarray}
\gamma_E &=& \pm \Delta_{\vec k_0} \alpha_{2 \vec k_0}
\label{condexp}
\end{eqnarray}
It is easy to check that at these points $E^{\pm}_{\vec k_0}=0$.

Thus, the perturbative Floquet theory predicts that the position of
both long-lived quasienergy modes and the presence/absence of
exceptional points can be tuned using the amplitude and frequency of
the drive. In the next section, we shall see that this statement
holds qualitatively for the exact spectrum. We note that the second
order Hamiltonian $H_{F \vec k}^{(2)}$ leads to smaller contribution
at large frequencies since its terms are suppressed by a overall
factor of $1/\omega_D$. However, its contribution to $H_{F \vec k}$
becomes important near special frequencies $\omega_m^{\ast}$ for
which $J_0[2g_1/(\hbar \omega_m^{\ast})]=0$. At these frequencies
$H_{F \vec k}^{(2)}$ contributes the only non-zero off-diagonal term
in $H_{F\vec k}$ (up to second order perturbation theory) and its
inclusion is therefore crucial in order to obtain a qualitative
match of the perturbative analytical results with exact numerics.

\subsection{Correlators, Entanglement and Fidelity}
\label{fptexp}

In this section, we shall express the correlation functions,
fidelity, and entanglement entropy of the driven integrable model in
terms of the eigenvalues and eigenvectors of $H_{F \vec k}$. This
will be particularly helpful in deducing their properties using the
expressions of second order Floquet energy derived in Eq.\
\ref{flham}.

We start by noting that the normalized eigenvectors of the second
order Floquet Hamiltonian corresponding to Floquet energies $E_{\vec
k}^{\pm}$ can be expressed in terms of components of a unit vector
$\vec n_{\vec k} = (n_{x \vec k},0, n_{z \vec k})$ where
\begin{eqnarray}
n_{x \vec k} &=& S_{2 \vec k}/E_{\vec k}, \quad n_{z \vec k}=
S_{1 \vec k}/E_{\vec k} \label{nunit}
\end{eqnarray}
In terms of these, the normalized eigenvectors of $H_{F \vec k}$
corresponding to quasienergies $\pm E_{\vec k}$ are given by
\begin{eqnarray}
|\pm ;\vec k\rangle &=& \frac{1}{{\mathcal N}_{\pm \vec k}}\left(
\begin{array}{c} p_{\pm \vec k} \\ q_{\pm \vec k}
\end{array} \right) \quad
p_{\pm \vec k} = n_{z \vec k} \pm 1, \nonumber\\
q_{\pm \vec k} &=& n_{x \vec k} \quad {\mathcal N}_{\pm \vec k} =
\sqrt{ |n_{z \vec k} \pm 1|^2 + |n_{x \vec k}|^2} \label{eigenvec}
\end{eqnarray}
Note that for $|g_0| \le 2$, $\Gamma_{\vec k}= {\rm Im}[E_{\vec k}]$
changes sign across $\vec k= \vec k^{\ast}$ for which $\alpha_{1 \vec
k^{\ast}}= -\alpha_{2 \vec k^{\ast}} \lambda \Delta^2_{\vec k^{\ast}}$. In
this case, for $\omega_D \simeq \omega_m^{\ast}$ where $S_{2 \vec
k}/E_{\vec k} \ll 1$ for all $\vec k$, and for $\gamma>0$, the
eigenfunction of $H_{F \vec k}$ corresponding to $\Gamma_{\vec k}>0$
changes from $\sim (0,1)^{T}$ to $\sim(1,0)^{T}$ sharply as one
crosses $\vec k^{\ast}$. In contrast, such a change is much more
gradual away from the special frequencies where $S_{2 \vec k}/E_{\vec
k}$ is not small.

In terms of $|\pm  ; \vec k\rangle$, it is possible to write the
evolution operator of the system at stroboscopic times $t_n= nT$ as
\begin{eqnarray}
U_{\vec k}(nT,0) &=& \sum_{a=\pm}  e^{-i a E_{\vec k} n T/\hbar}
|a;\vec k\rangle\langle a; \vec k| \label{evolop}
\end{eqnarray}
We note that the evolution operator $U_{\vec k}(nT,0)$ is not
unitary due to non-zero $\Gamma_{\vec k}$. Consequently, to obtain
the state after $n$ drive cycles, we need to adapt the standard
normalization procedure for non-Hermitian systems
\cite{nhdyn1,nhdyn2,nhdyn3,nhdyn4,nhdyn5} which yields
\begin{eqnarray}
|\psi_{\vec k}( nT)\rangle &=& \frac{|\tilde \psi_{\vec k}(
nT)\rangle}{|\langle \tilde \psi_{\vec k}(nT)|\tilde \psi_{\vec k}(
nT)\rangle|},\nonumber\\
|\tilde \psi_{\vec k}(nT)\rangle &=&  U_{\vec k}(nT,0) |\psi_{0\vec
k}\rangle \label{normfn}
\end{eqnarray}
where $|\psi_{0\vec k}\rangle$ is the initial state. In what
follows, we shall parameterize the initial state $|\psi_{0\vec
k}\rangle= (u_{0\vec k}, v_{0\vec k})^T$ using an angle $\theta_{0
\vec k}$ such that $u_{0\vec k}= \cos \theta_{0\vec k}$ and $v_{0
\vec k}=\sin \theta_{0 \vec k}$. This allows us to write, using
Eqs.\ \ref{eigenvec}, \ref{evolop}, and \ref{normfn}, $|\psi_{\vec
k} (nT)\rangle = (u_{\vec k}(nT), v_{\vec k}(nT))^T$ where
\begin{eqnarray}
u_{\vec k}(nT) &=& \frac{ \sum_{a=\pm} e^{-i a E_{\vec k} n T/\hbar}
\mu_{a \vec k} p_{a \vec k}}{{\mathcal D}_{\vec k}(\theta_{0 \vec k})} \nonumber \\
v_{\vec k}(nT) &=&  \frac{ \sum_{a=\pm} e^{-i a E_{\vec k} n
T/\hbar} \mu_{a \vec k} q_{a \vec
k}}{{\mathcal D}_{\vec k}(\theta_{0 \vec k})} \label{uveq} \\
{\mathcal D}_{\vec k}(\theta_{0 \vec k}) &=& \Big[|\sum_{a=\pm}
e^{-i E_{a \vec k}n
T/\hbar} p_{a \vec k} \mu_{a \vec k}|^2 \nonumber\\
&& + |\sum_{a=\pm} e^{-i E_{a \vec k} n T/\hbar} q_{a \vec k}
\mu_{a \vec k}|^2 \Big]^{1/2} \nonumber\\
\mu_{\pm \vec k} &=&  p_{\pm \vec k}^{\ast} \cos \theta_{0 \vec
k} + q^{\ast}_{\pm \vec k} \sin \theta_{0 \vec k},\nonumber
\end{eqnarray}
Using this wavefunction, one can define the fidelity $\chi(nT)
=\prod_{\vec k} \chi_{\vec k}(nT)$ where $\chi_{\vec k}(nT) =
|\langle \psi_{0 \vec k}|\psi_{\vec k}(nT)\rangle|^2$. In what
follows, we shall be mainly interested in studying the behavior of
$g(nT)= \ln \chi(nT)\;$ \cite{fidref1}. Using Eqs.\ \ref{normfn}
and \ref{uveq}, one can express $g(nT)$ as
\begin{eqnarray}
g(nT) &=& \int \frac{d^d k}{V_0} \ln |u_{\vec k}(nT) \cos \theta_{0
\vec k} + v_{\vec k}(nT) \sin \theta_{0 \vec k}|^2 \nonumber
\\ \label{fideq}
\end{eqnarray}
where $V_0=(2 \pi)^d/2$ denotes the volume of the $d$-dimensional
Brillouin zone.

Next, we compute the correlation functions of the model. For the
class of integrable models discussed here, the non-trivial
correlation functions are given by
\begin{eqnarray}
N_{\vec k}(nT) &=& \langle (2 c_{\vec k}^{\dagger} c_{\vec
k}-1)\rangle= 2 |v_{\vec k}(nT)|^2-1  \label{correxp1} \\
F_{\vec k} (nT) &=& \langle c_{\vec k} c_{-\vec k} +{\rm h.c.}
\rangle = (u_{\vec k}^{\ast}(nT) v_{\vec k}(nT) +{\rm h.c.} )
\nonumber
\end{eqnarray}
The real space correlation functions can be obtained via Fourier
transforms of $N_{\vec k}(nT)$ and $F_{\vec k}(nT)$.

Finally we note that for this class of integrable models the
entanglement entropy can be expressed in terms of the correlation
matrix ${\mathcal C}$. For a 1D fermionic chain of length $L$ and a
subsystem of size $\ell \le L$, the correlation matrix can be
written as \cite{nhdyn5}
\begin{widetext}
\begin{eqnarray}
{\mathcal C} &=& \left( \begin{array}{cccc} \Pi_0 & \Pi_{-1} & ..
& \Pi_{1-\ell} \\ \Pi_{1} & \Pi_0 & .. & \Pi_{2-\ell} \\
.. & .. & .. & .. \\
\Pi_{\ell-1} & \Pi_{\ell-2} & .. & \Pi_0 \end{array} \right), \quad
\Pi_{\ell_0} = \int \frac{d^d k}{V_0} e^{ i k  \ell_0} \hat \Pi_{k}
\cdot \hat \tau_k \nonumber\\
\Pi_{y k} &=&  2|v_{k}(nT)|^2-1, \quad  \Pi_{x k} = 2\;{\rm Re}
(u_{k} (nT) v_{k}^{\ast} (nT)), \quad \Pi_{z k}= 2 \;{\rm Im}
(u_k (nT) v_k^{\ast} (nT)) \label{entangdef}
\end{eqnarray}
\end{widetext}
The entanglement entropy can then be computed using eigenvalues
$\zeta_r(nT)$, where $r=1\,..\,2\ell$, of ${\mathcal C}$. In terms
of these one obtain the von-Neumann entropy as
\begin{eqnarray}
S_{\ell}(nT)  &=& -\sum_{r=1}^{2 \ell} \zeta_r (nT)\ln \zeta_r(nT)
\label{entangexp}
\end{eqnarray}
We shall use these expressions to compute the correlations, fidelity
and entanglement both from exact numerics and using the second order
perturbative Floquet Hamiltonian for the 1D Ising
chain in the next section.

\section{Numerical results} \label{numerics}

In this section, we present our numerical results for the driven,
non-Hermitian 1D Ising chain with the Hamiltonian given by Eq.\
\ref{isham}. Using a standard Jordan-Wigner transformation
\cite{subir1}, the Ising chain (Eq.\ \ref{isham}) can be mapped into
the free fermion Hamiltonian (Eq.\ \ref{hamdiv}) with the
identification $a_{3 k}= 2\cos k$, $J=1$, $g(t)= 2 h(t)= 2(h_0+h_1
\cos \omega_D t)$, and $\Delta_{k}=2\sin k$. In this notation, the
ferromagnetic state with spin-up on all sites is mapped to fermion
vacuum. We note that such a transformation provides a direction
relation between the fermion density operator $\hat n_j =
c_j^{\dagger} c_j$ (where $c_j$ denotes the fermion annihilation
operator on site $j$) and $\sigma_j^z$ as $\sigma_j^z= 1-2 \hat
n_j$.

We present our results obtained using both exact numerical
computation of $U_k(T,0)$ and using $H_{F k}$ (Eq.\ \ref{flham})
computed using second order FPT. For the former, we follow the
standard procedure of Suzuki-Trotter decomposition of $U_k$ into
$n_0$ steps of width $\delta t = T/n_0$. The width of these time
steps are chosen such that $H_k(t)$ (Eq.\ \ref{hamdiv}) does not
change significantly within each of these steps.  This allows one to
numerically compute the evolution operator as
\begin{eqnarray}
U_k(T,0) &=& \prod_{j=1,n_0} U_k(t_j, t_{j-1})= \prod_{j=1,n_0}
e^{-i \delta t H_k(t_j)/\hbar} \nonumber\\ \label{unumerics}
\end{eqnarray}
One can then diagonalize $U_k$ to find out its eigenvalues $e^{ i
\theta_k^{\pm}}$ (where $\theta_k^{\pm}= E_{k}^{\pm} T/\hbar$ are in
general complex numbers) and the corresponding eigenvectors
$|\pm;k\rangle$. This leads to the evolution
operator
\begin{eqnarray}
U_k (nT,0) &=& \sum_{a=\pm} e^{-i n \theta_k^a} |a;k\rangle \langle
a;k|. \label{evolnum}
\end{eqnarray}
Using Eq.\ \ref{evolnum} one can compute fidelity, correlation
function and entanglement entropy numerically following the steps
outlined in Sec.\ \ref{fptexp}

\subsection{Floquet spectrum}
\label{fspec}

\begin{figure}
\rotatebox{0}{\includegraphics*[width= 0.49 \linewidth]{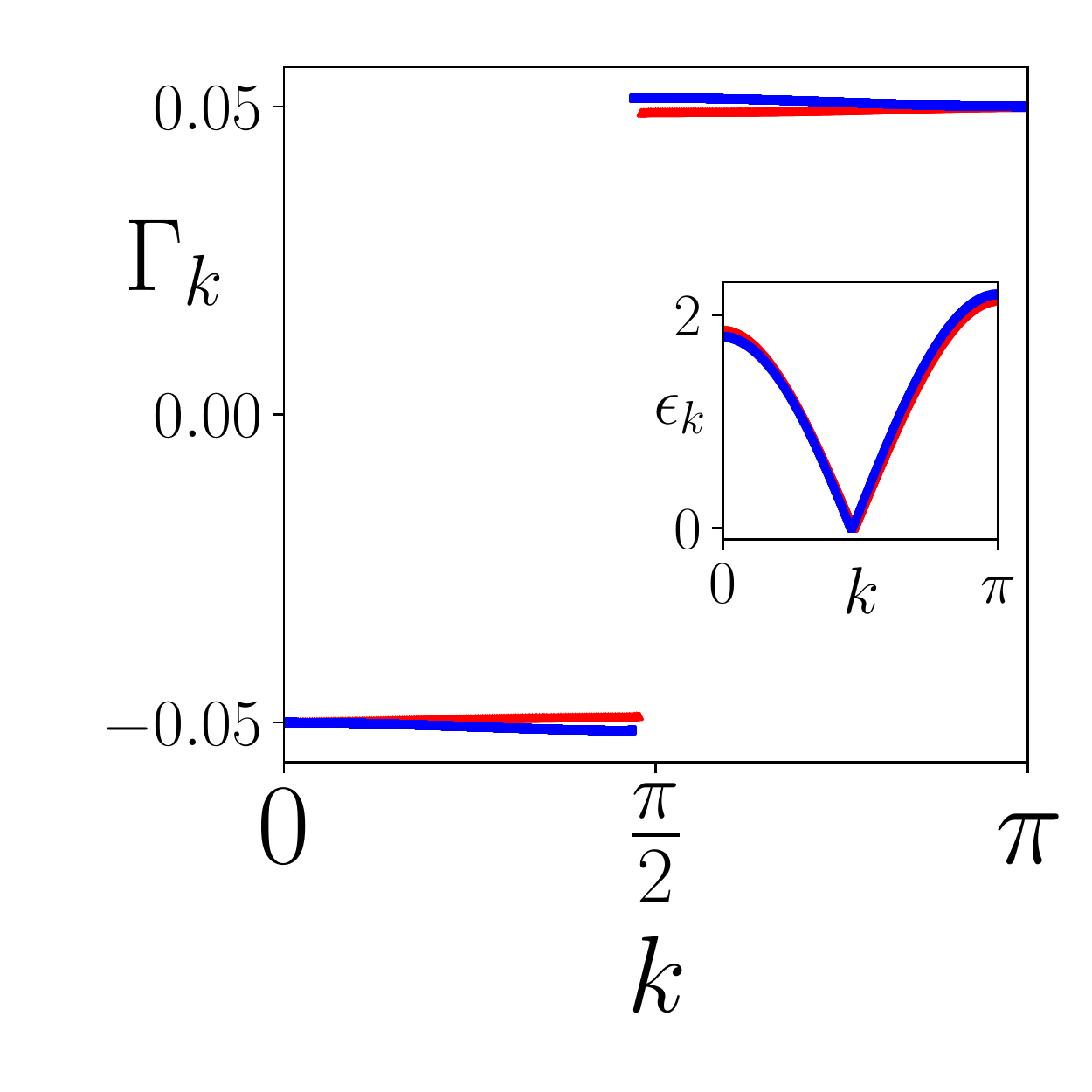}}
\rotatebox{0}{\includegraphics*[width= 0.49 \linewidth]{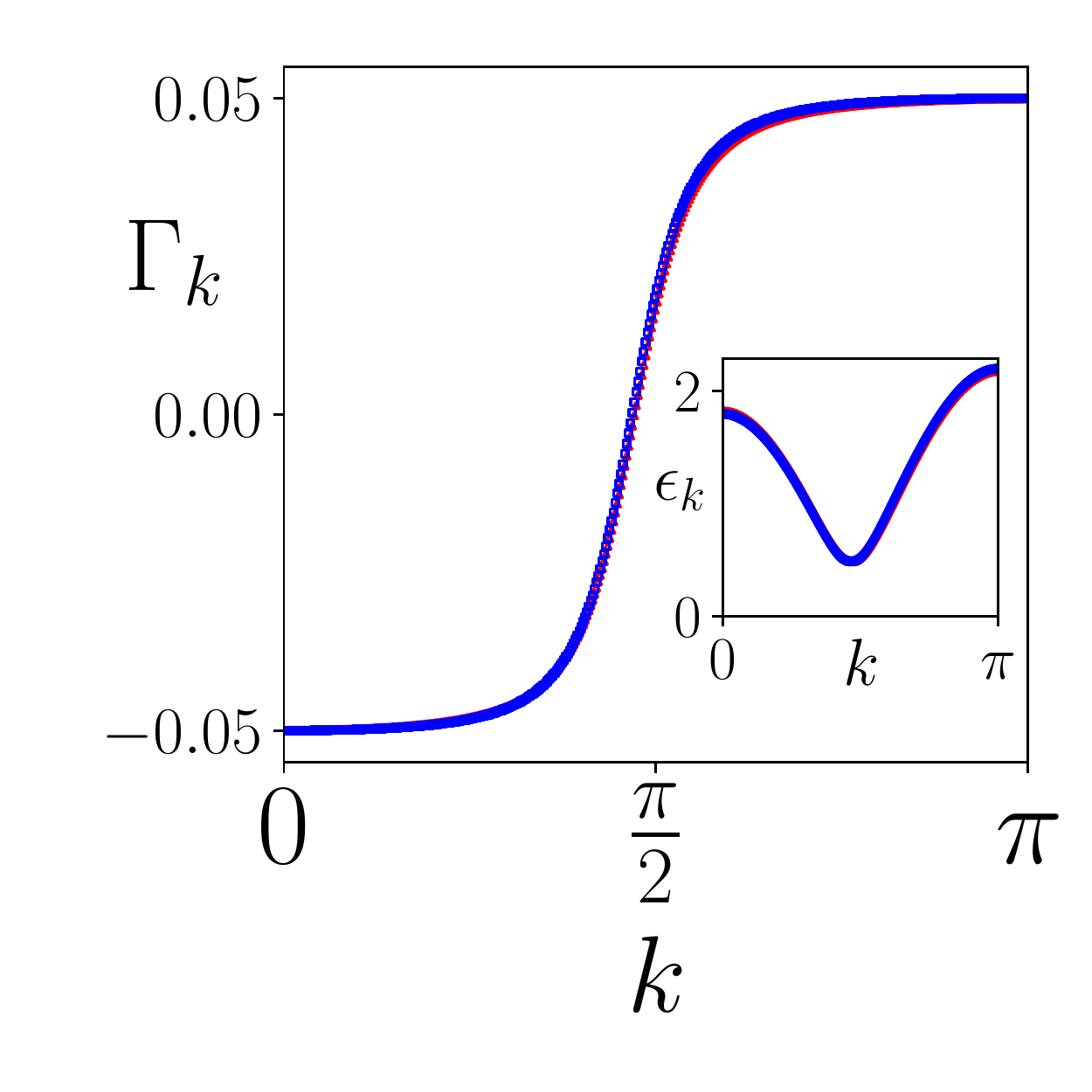}}
\rotatebox{0}{\includegraphics*[width= 0.49 \linewidth]{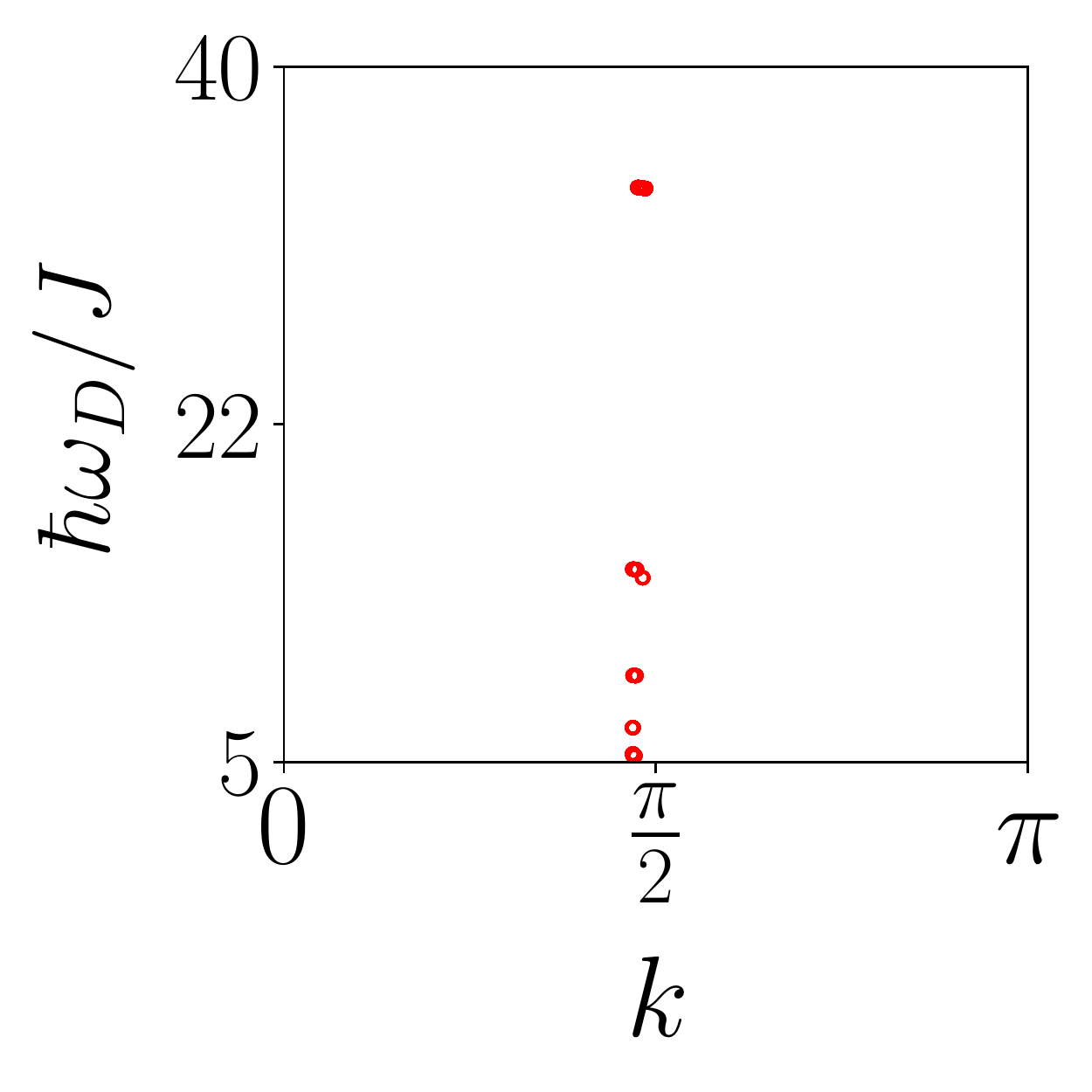}}
\rotatebox{0}{\includegraphics*[width= 0.49 \linewidth]{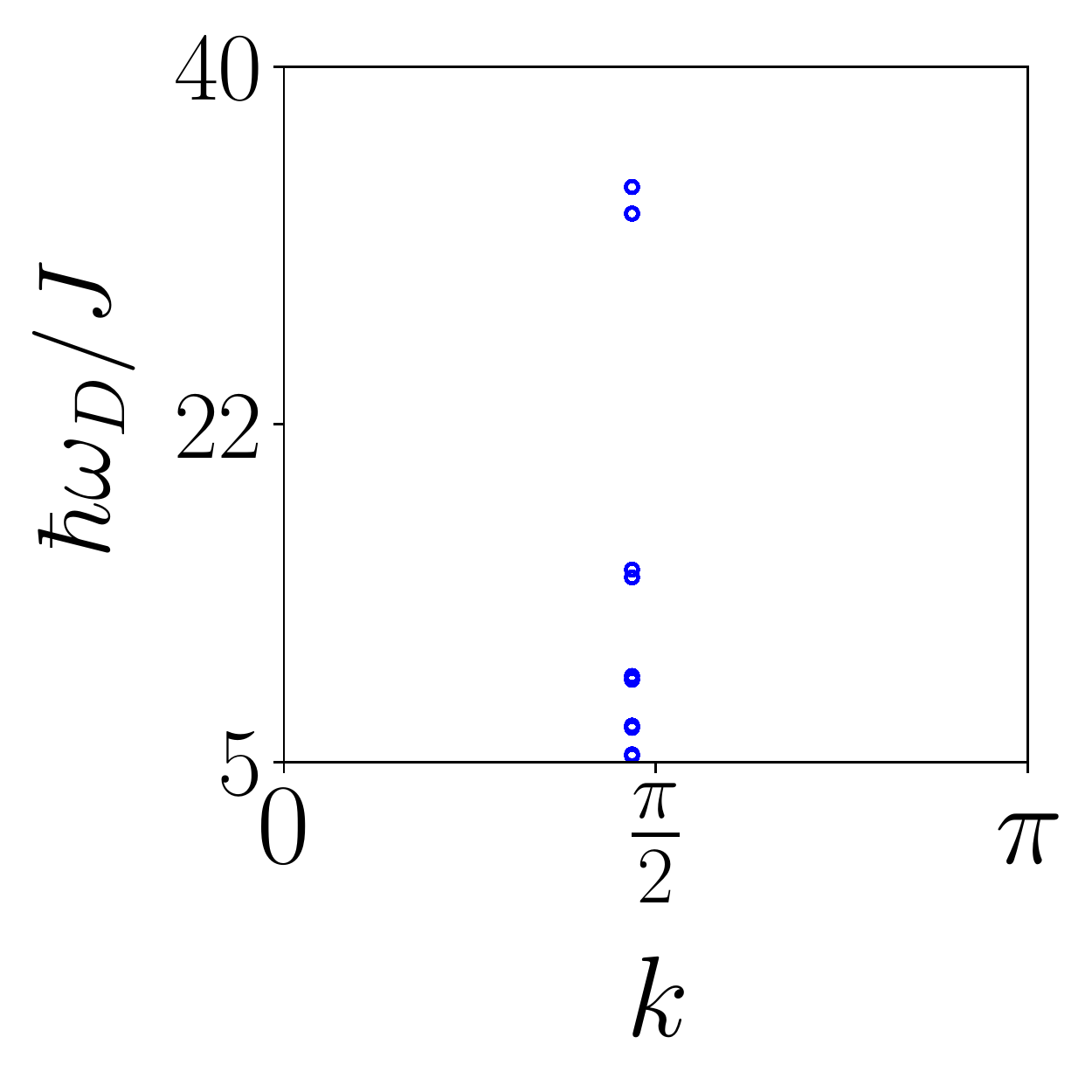}}
\caption{Top Left Panel: Plot of $\Gamma_k$ as a function of $k$ for
the branch with $\epsilon_k>0$ and with $\hbar \omega_D/J=9.24$. The
inset shows $\epsilon_k$ as a function of $k$. Top right panel:
Similar plot for $\hbar \omega_D/J=8$. The red(blue) lines
represents results obtained using exact numerics with system size
$L=1000$ (second order FPT). Bottom left panel: Plots of exceptional
points as a function of $\hbar \omega_D/J$ and $k$ as obtained using
exact numerics. Bottom right panel: Same as the bottom left panel as
obtained from second order FPT by solving Eq.\ \ref{condexp}. For
all plots $h_0=0.1,\gamma=0.05$, $h_1=20$ and energy scales are
measured in units of $J$. See text for details.} \label{fig1}
\end{figure}

In this subsection we present our results for the Floquet spectrum.
To this end, we plot $\Gamma_k$ and $\epsilon_k$ (Eq.\ \ref{fldisp})
in top panels of Fig.\ \ref{fig1} for $\hbar \omega_D/J=9.24$(top
left panel) and $8$ (top right panel). The branch of $E_k^{\pm}$
with $\epsilon_k>0$ is plotted in Fig.\ \ref{fig1}. Both the figures
show a change in sign of $\Gamma_k$ around $k=k^{\ast}\sim 1.5$. The
value of $k^{\ast}$ is consistent with that found from the condition
$k^{\ast} \simeq  \arccos h_0$; this is due to the fact that the
second order contributions to the Floquet spectrum are small
compared to the first order terms. This shows that such a change in
sign of $\Gamma_k$ is contingent on the condition $|h_0|\le 1$. The
change of sign is gradual away from $\omega_m^{\ast}$ as shown for
$\hbar \omega_D/J=8$ in the right panel; in contrast it is abrupt
for $\hbar \omega_D/J=9.24$ which corresponds to $\omega_D=
\omega_3^{\ast}$. We have checked that a similar behavior holds near
all other $\omega_D=\omega_m^{\ast}$. We also note that the second
order FPT (blue lines) shows an excellent match with the exact
results (red lines) for all $k$.

The corresponding bottom panels shows the position of the
exceptional points as a function of $\hbar \omega_D/J$ and $k$ where
both real and imaginary components of $E_k$ vanishes. The bottom
left panel of Fig.\ \ref{fig1} shows the positions of the
exceptional points obtained from exact numerics; this is determined
numerically by choosing $|E_k| \le \delta$ where $\delta \sim
10^{-2}$. We have checked that lowering $\epsilon$ further does not
change the nature of the plots. The bottom right panel shows similar
points obtained from second order FPT by solving Eq.\ \ref{condexp}.
We note that near the special frequencies where $J_0(4h_1/(\hbar
\omega_D))=0$ leading to very small off-diagonal terms, our choice
of parameters do not allow for exceptional point; this is clearly
seen in the bottom panels of Fig.\ \ref{fig1}.  Moreover such points
form discrete set of points in $k$ space; consequently their
presence do significantly affect the dynamics of magnetization or
correlation functions which involves sum over all $k$ points.

Before concluding this section, we note that the Floquet spectrum
obtained from the second order FPT matches quite well with exact
numerics; moreover, the position of the exceptional points in the
$\omega_D-k$ plane obtained by exact numerics also matches that
obtained from Eq.\ \ref{condexp}. Thus these results confirm the
validity of second order FPT for a wide range of $\omega_D$.

\subsection{Fidelity and Correlations}
\label{fidcor}

\begin{figure}
\rotatebox{0}{\includegraphics*[width= 0.49 \linewidth]{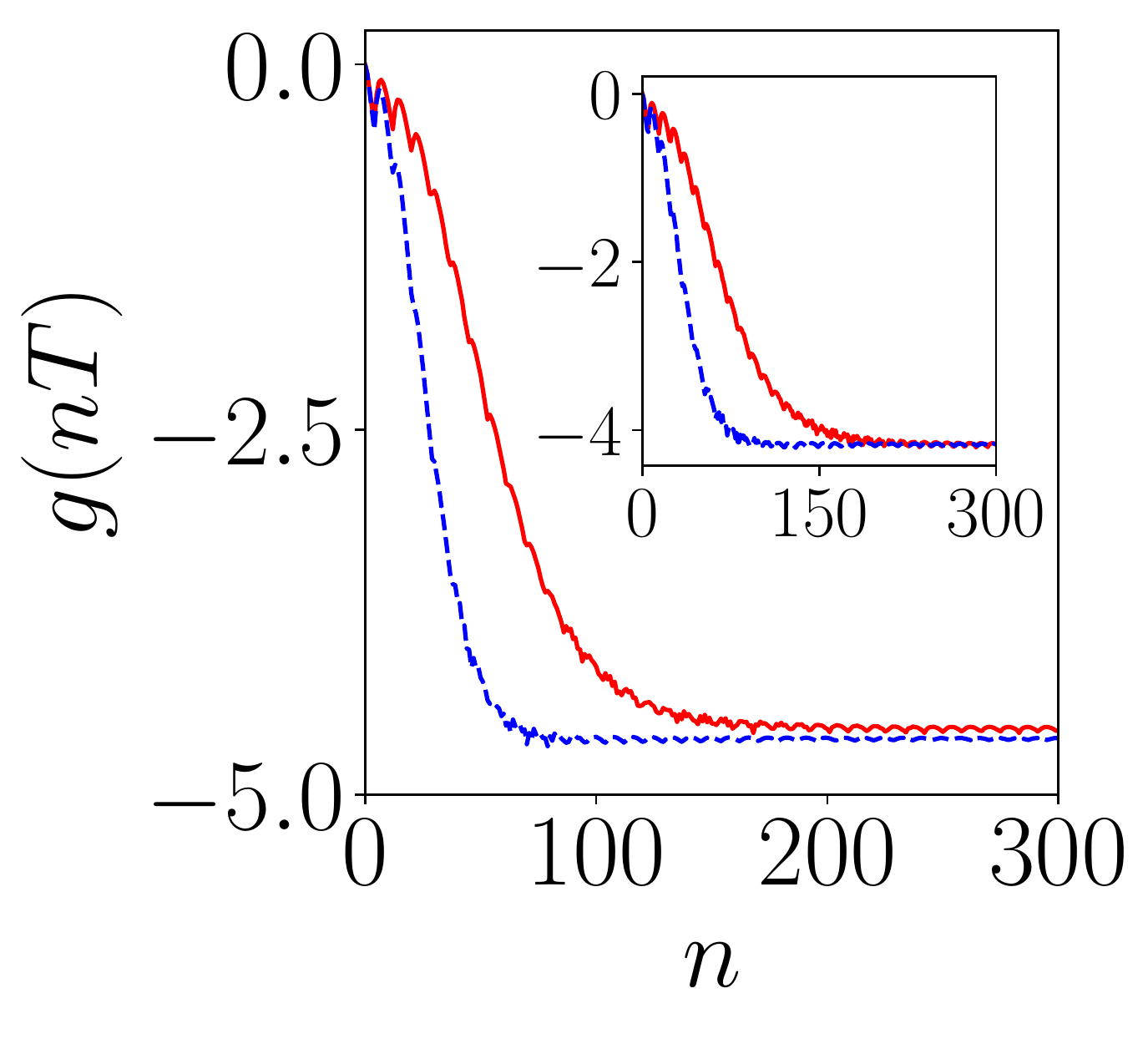}}
\rotatebox{0}{\includegraphics*[width= 0.49 \linewidth]{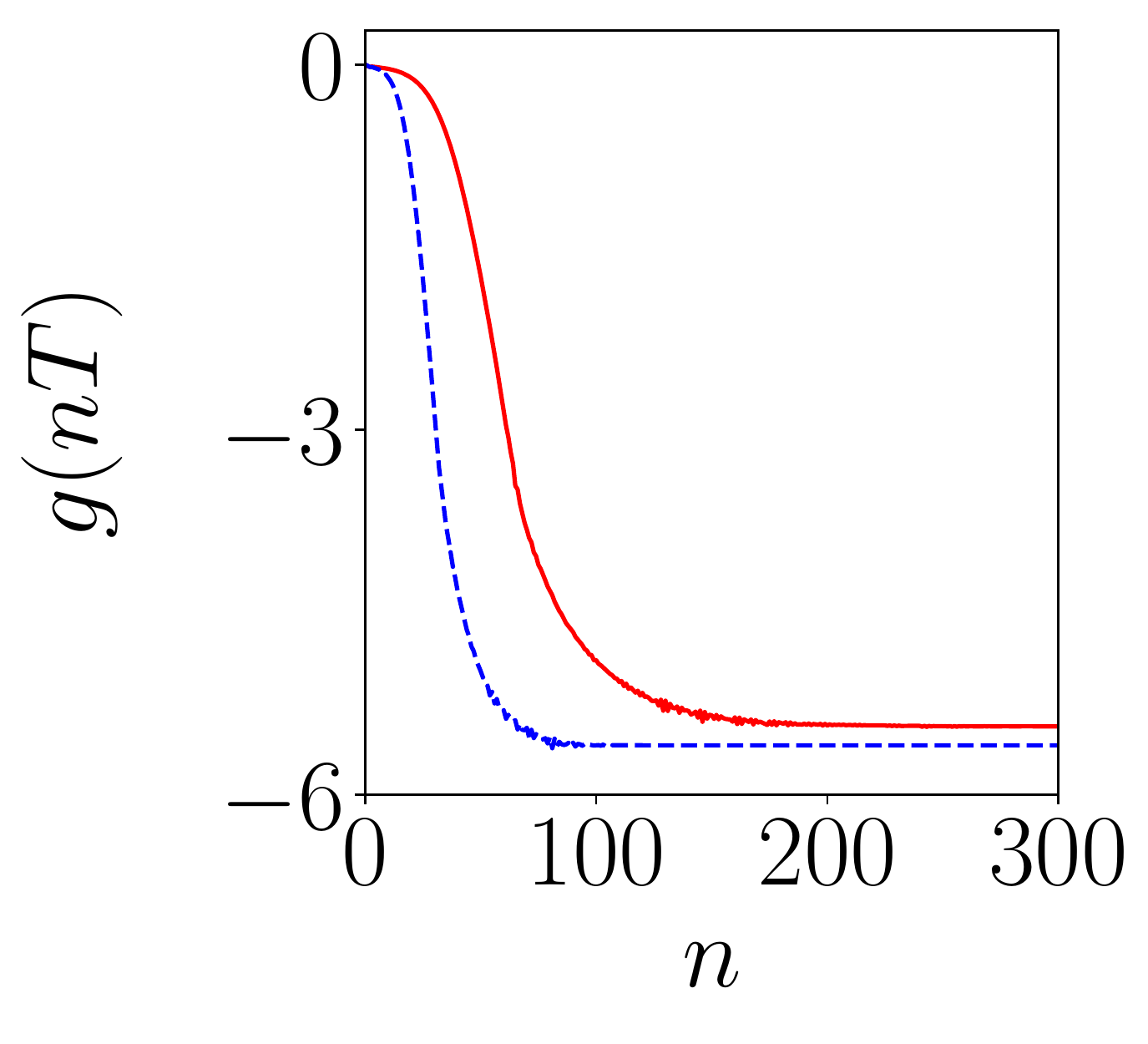}}
\rotatebox{0}{\includegraphics*[width= 0.49 \linewidth]{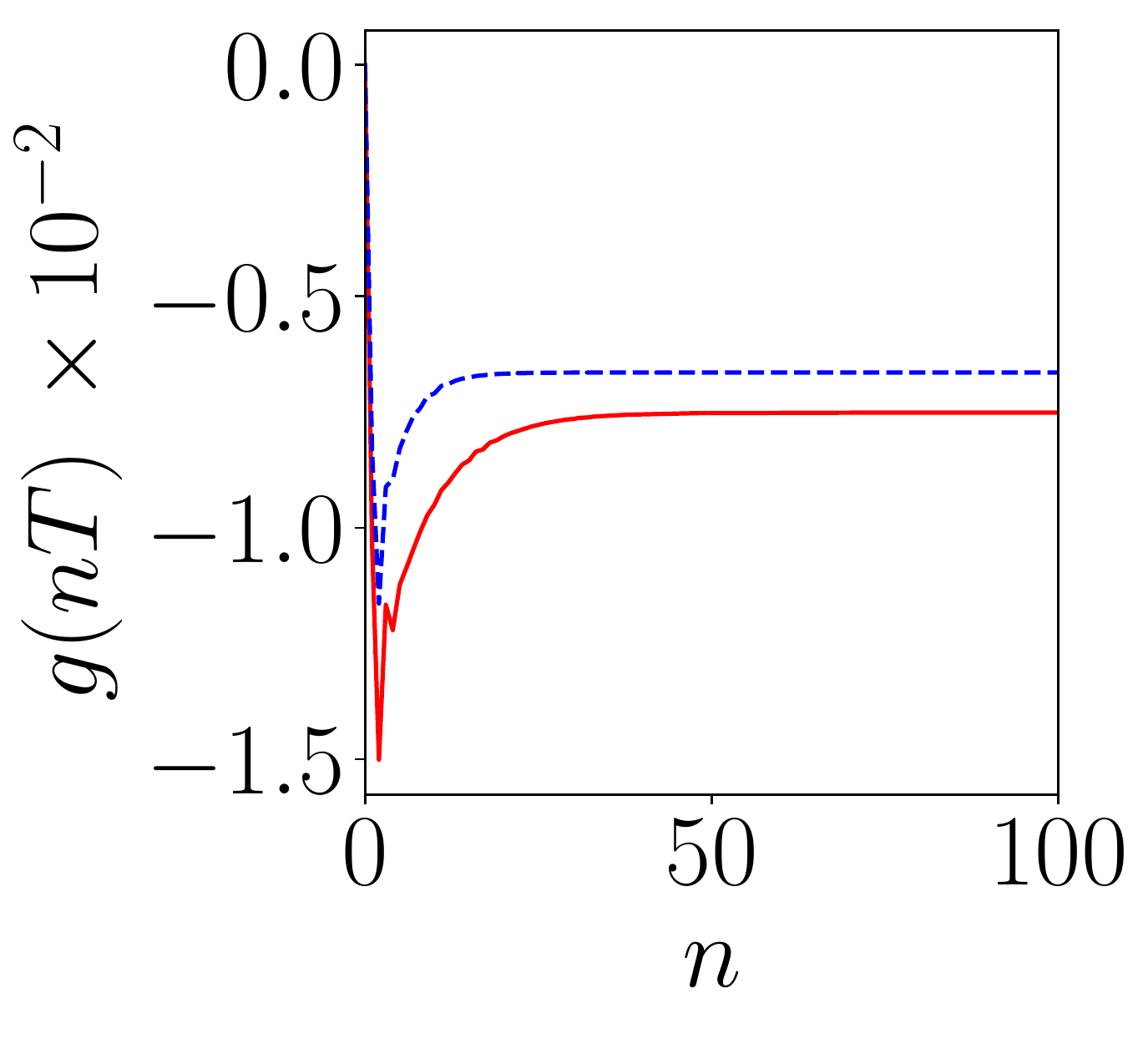}}
\rotatebox{0}{\includegraphics*[width= 0.49 \linewidth]{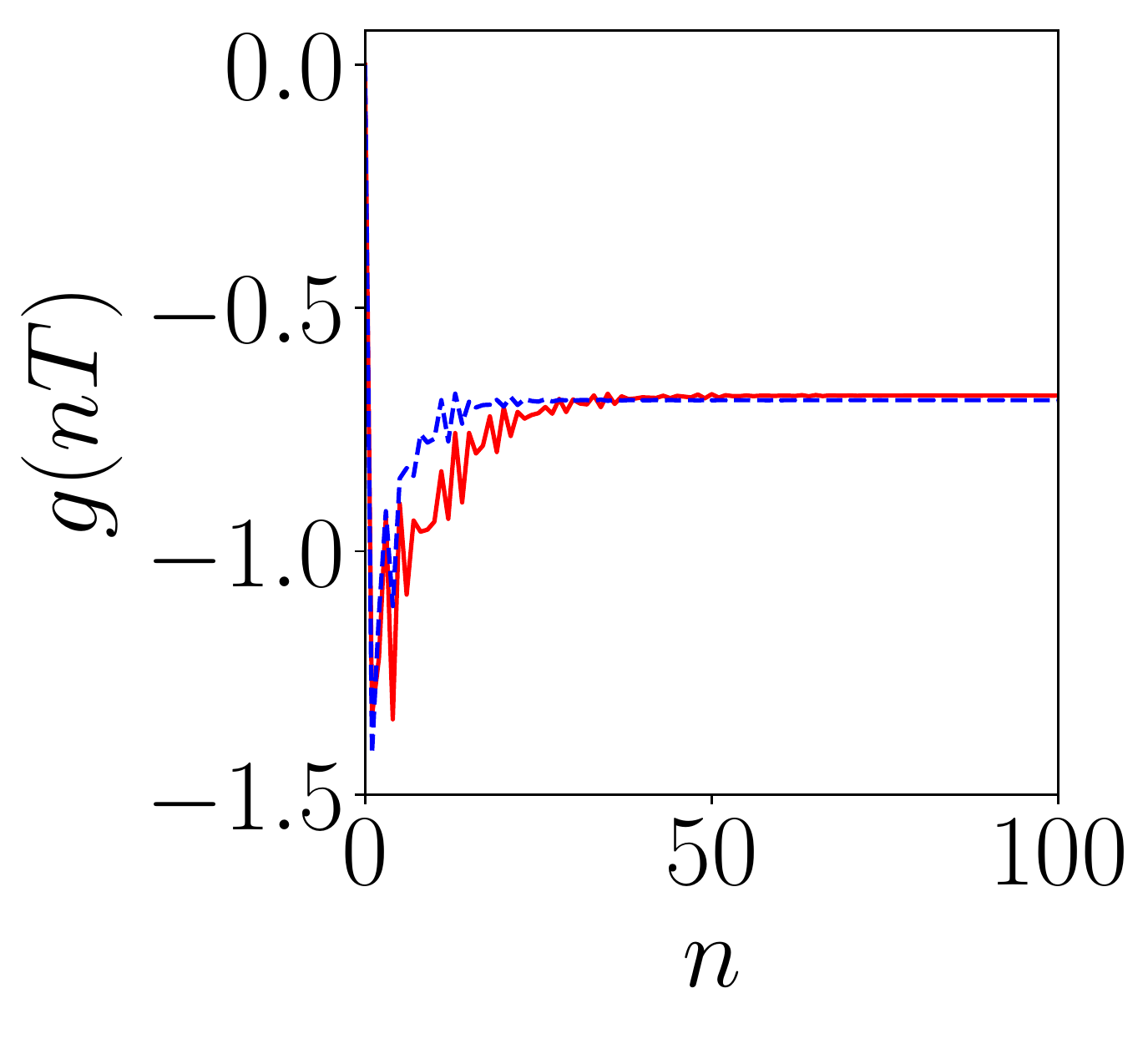}}
\caption{Top Left Panel: Plot of $g(nT)$ as a function of $n$ for
$\hbar \omega_D/J=8$ and $|\psi_0\rangle= \prod_k (0,1)^{T}$ which
corresponds to the all spin-down ferromagnetic initial state. The
inset shows analogous plot for $\hbar \omega_D/J=11$. Top right
panel: Similar plot for $\hbar \omega_D/J=9.24$. Bottom left panel:
Similar plot for $\hbar \omega_D/J=9.24$ with the initial state
$|\psi_0\rangle= \prod_k (1,0)^{T}$ which corresponds to the all
spin-up ferromagnetic state. Bottom right panel: Similar plot for
$\hbar \omega_D/J=9.24$ with the initial state $|\psi_0\rangle=
\prod_k (1,1)^{T}/\sqrt{2}$. For all plots red(blue) represents
results obtained from exact numerics (second order FPT). All other
parameters are same as in Fig.\ \ref{fig1}. See text for details.}
\label{fig2}
\end{figure}

In this section, we first study the fidelity $g(nT)$ (Eq.\
\ref{fideq}) of the driven model as a function of $n$ for several
representative values of $\omega_D$. These plots for shown in the
top panels of Fig.\ \ref{fig2} for an initial state
$|\psi_{0}\rangle= \prod_k (u_{0k},v_{0k})^T= \prod_k (0,1)^T$ while
the bottom panel shows analogous plots for $|\psi_{0}\rangle=\prod_k
(1,0)^T$ (bottom left panel) and $|\psi_{0}\rangle=\prod_k
(1,1)^T/\sqrt{2}$ (bottom right panel).

The top left panel of Fig.\ \ref{fig2} shows the behavior of $g(nT)$
for $\hbar \omega_D/J=8$ and $\hbar \omega_D/J=11$
(inset). These frequencies are far away from $\omega_m^{\ast}$ (for
$m=1,2,3 ...$) for which $J_0(4h_1/(\hbar \omega_D))=0$; thus the
behavior of $g(nT)$ in this plot represent its typical behavior for
a ferromagnetic initial state at most frequencies. The plot
indicates a decay of $g(nT)$ to its steady state value with small
but finite oscillations. These features are predicted by both second
order FPT (blue dashed lines) and exact numerics (red solid lines);
the perturbative prediction match the exact results quite well at
these frequencies.

In contrast, the top right and the bottom panels show the behavior
of $g(nT)$ for $\hbar \omega_D/J=9.24$ which corresponds to
$\omega_D=\omega_3^{\ast}$. The top right panels show lack of
oscillations along with a steady state value of $g(nT) \ll 0$. In
contrast, the plot of $g(nT)$ in the bottom left panel, which
corresponds to an initial state $\prod_k (1,0)^T$, yields a
near-zero steady state value. This indicates a high overlap of the
steady state with the initial state. The bottom right panel,
corresponding to $|\psi_0 \rangle=\prod_k (1,1)^T/\sqrt{2}$, shows
oscillatory nature of $g(nT)$ along with a steady state value of
$\sim \ln (1/2) $.

\begin{figure}
\rotatebox{0}{\includegraphics*[width= 0.49 \linewidth]{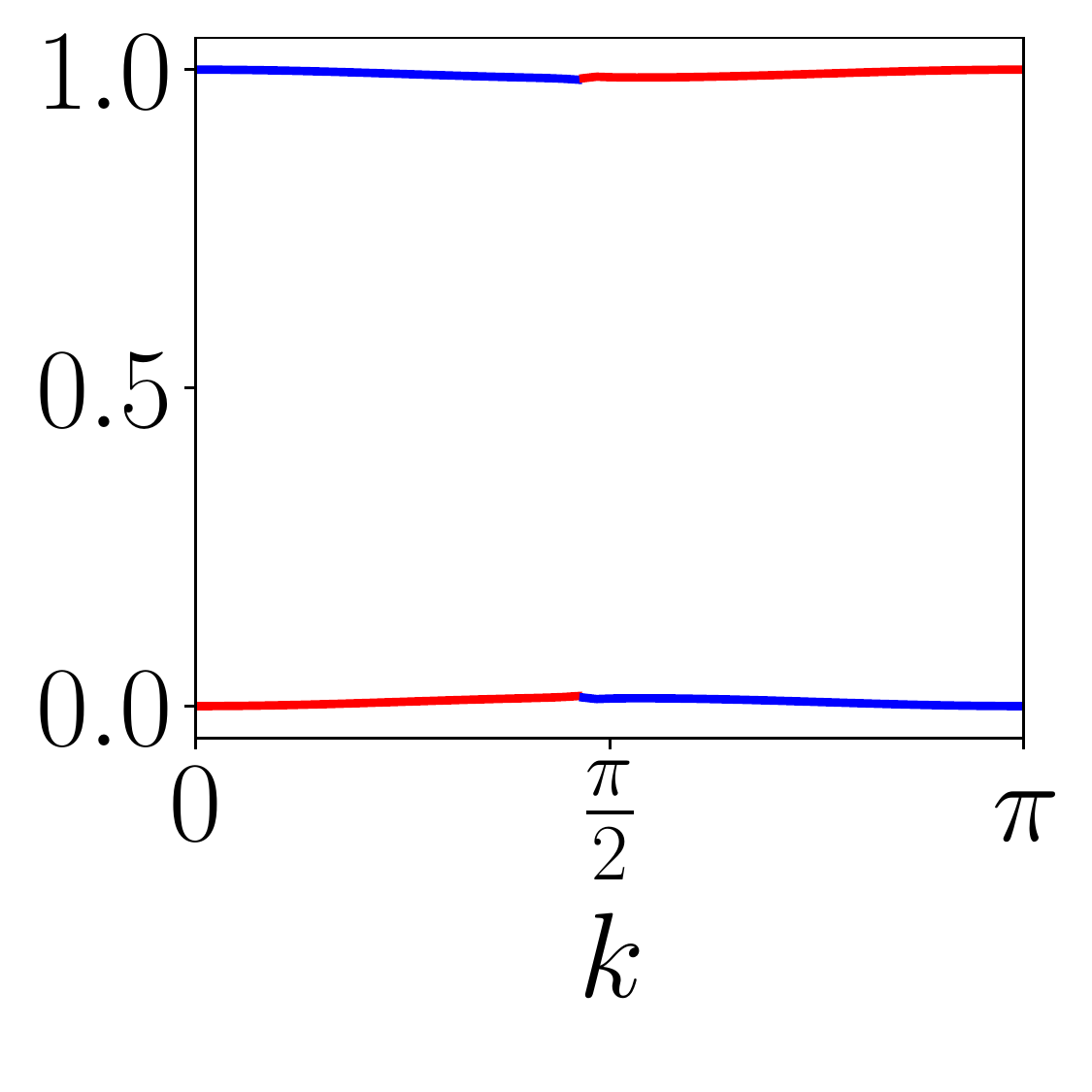}}
\rotatebox{0}{\includegraphics*[width= 0.49 \linewidth]{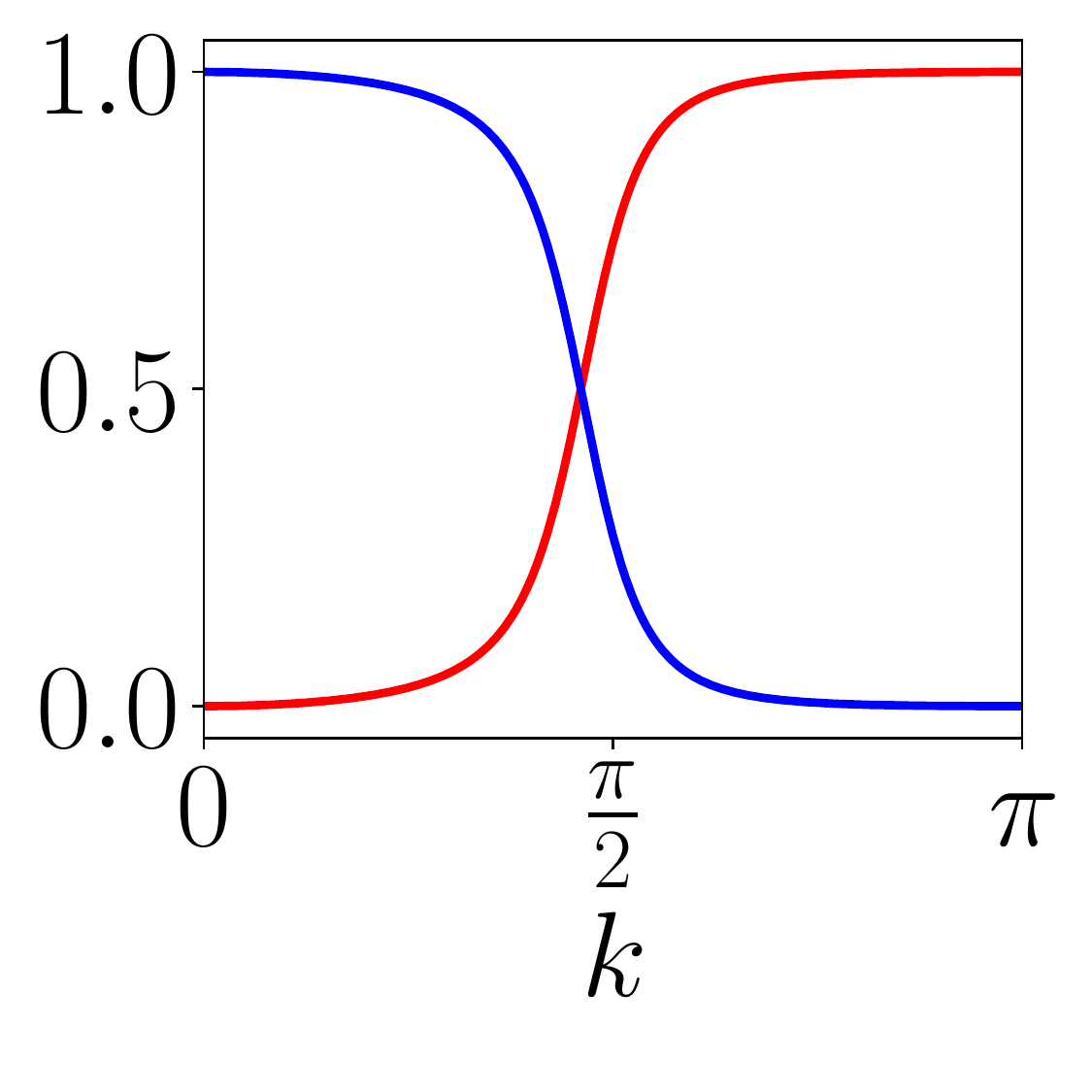}}
\caption{Plot of $|p_{+k}|^{2}$ (red solid line) and $|p_{- k}|^{2}$
(blue solid line), obtained using second order FPT, as a function
$k$ for $\hbar \omega_D/J=9.24$ (left panel) and $\hbar
\omega_D/J=8$ (right panel). The behavior $q_{\pm k}$ can be read
off from these plots using the relation $|p_{\pm k}|^2 + |q_{\pm
k}|^2=1$. All other parameters are same as in Fig.\ \ref{fig1}. See
text for details.} \label{fig3}
\end{figure}

To qualitatively understand these features, we first consider the
initial state $\prod_k (0,1)^T$. For any given $k$, the wavefunction
overlap for this initial state is given by $\chi_k(nT) =
|v_k(nT)|^2$. Moreover, for this initial state, $\mu_{\pm k} =
q^{\ast}_{\pm k}$ (Eq.\ \ref{uveq}). This leads to
\begin{eqnarray}
\chi_k(nT) &=& \frac{| |q_{+ k}|^2 e^{-i E_k nT/\hbar} + |q_{- k}|^2
e^{i E_k nT/\hbar}|^2}{|{\mathcal D}_{k}(\pi/2)|^2}  \nonumber\\
&=& \frac{1}{\mathcal{D}_k^2(\pi/2)} \Big( |q_{+k}|^4 e^{2 \Gamma_k
nT/\hbar} + |q_{-k}|^4 e^{-2\Gamma_k nT/\hbar} \nonumber\\
&& +2 |q_{+ k} q_{- k}|^2 \cos (2 \epsilon_k nT/\hbar) \Big)
\label{vkeq}
\end{eqnarray}
where we have used $E_k = \epsilon_k + i \Gamma_k$ and ${\mathcal
D}_{k}(\pi/2) = {\mathcal D}_k (\theta_{0k}=\pi/2)$ (Eq.\
\ref{uveq}). The first two terms in the expression of $\chi_k$
determines its steady state behavior while the last term yields the
intermediate oscillation.

The nature of the fidelity can be qualitatively understood from the
behavior of $q_{\pm k}$ and $\Gamma_k$ as a function of $k$. The
plots of $|p_{\pm k}|^2=1-|q_{\pm k}|^2$ is shown in Fig.\
\ref{fig3} as a function of $k$. From this plot, we note that near
the special frequencies $\omega_m^{\ast}$, $|q_{- k}|^2 \sim
\theta(k-k^{\ast})$ and $|q_{+ k}|^2 \sim \theta(k^{\ast}-k)$; thus
the oscillations in $\chi_k(nT)$ whose amplitude $\sim |q_{+k}
q_{-k}|^2$ vanishes for all $n$. Furthermore when $ nT
|\Gamma_k|/\hbar \ll 1$,  such that ${\mathcal D}_k(\pi/2) \sim
(|q_{+ k}|^2 + |q_{- k}|^2)^{1/2}$, we find
\begin{eqnarray}
\chi_k \sim |q_{- k}|^2 \theta(k-k^{\ast}) + |q_{+ k}|^2
\theta(k^{\ast}-k) \simeq 1.
\label{01state_fidelity}
\end{eqnarray}
In contrast, for large $n$, where $ nT |\Gamma_k|/\hbar \gg 1$
 one has
\begin{eqnarray}
{\mathcal D}_{k}^2(\pi/2) &\simeq& |q_{+k}|^2 \exp\left(2
\Gamma_k nT/\hbar\right) \nonumber\\
&& + |q_{-k}|^2 \exp\left(-2 \Gamma_k n T/\hbar\right),
\label{dkexp}
\end{eqnarray}
the expression of $\chi_k$ can be written as
\begin{eqnarray}
\chi_k &\sim& |q_{-k}|^2 \left[ 1+ \left(\frac{|q_{+ k}|
e^{-|\Gamma_k| nT/\hbar}}{|q_{-k}|} \right)^4 \right] \quad k<
k^{\ast}
\nonumber\\
&\sim & |q_{+ k}|^2 \left[ 1+ \left(\frac{|q_{- k}| e^{- |\Gamma_k|
nT/\hbar}}{|q_{+k}|} \right)^4 \right] , \quad k > k^{\ast}
\nonumber\\ \label{01state}
\end{eqnarray}
where we have used the fact that $\Gamma_k <(>) 0$ for
$k<(>)k^{\ast}$ as shown in the top panels of Fig.\ \ref{fig1}. This
shows that $g(nT)$ (Eq.\ \ref{fideq}) assumes a large negative value
at large $n$; moreover, the decay to the steady state is
exponential. The steady state value of $g$ depends on
$\ln \chi_k$ for $ nT |\Gamma_k|/\hbar \gg 1$; near
$\omega=\omega_m^{\ast}$ where $q_{\pm k}$ shows a sharp jump around
$k=k^{\ast}$, $\chi_k \sim |q_{\pm k}|^2 \to 0$ for all $k$ as
can be seen from Fig.\ \ref{fig3}. Thus $g \sim \int dk \ln \chi_k$
assumes a large negative value as can be seen from the top right
panel of Fig.\ \ref{fig2}.

For the initial state $\prod_k (1,0)^T$, we find that $ \mu_{\pm k}
= p_{\pm k}^{\ast}$. Using this, a similar calculation yields
\begin{eqnarray}
\chi_k(nT) &=& \frac{1} {\mathcal{D}_k^2(0)} \Big( |p_{+ k}|^4 e^{2
\Gamma_k nT/\hbar} + |p_{- k}|^4 e^{-2\Gamma_k nT/\hbar}
\nonumber\\
&& + 2|p_{+k}p_{- k}|^2 \cos\left( 2\epsilon_k nT/\hbar\right) \Big)
\label{vk1eq}
\end{eqnarray}
We note that near $\omega_m^{\ast}$ for $|\Gamma_k| nT/\hbar \ll 1$,
a similar analysis as given in Eq.\ \ref{01state_fidelity} yields
$\chi_k \sim 1$ . In contrast for the steady state where $|\Gamma_k|
nT/\hbar \gg 1$, we find
\begin{eqnarray}
\chi_k &\sim& |p_{- k}|^2 \left( 1 + \frac{|p_{+k}|^4
\exp\left(-4|\Gamma_k|nT/\hbar\right)}{|p_{-k}|^4} \right) \quad k<
k^{\ast}
\nonumber\\
&\sim & |p_{+ k}|^2 \left( 1 + \frac{|p_{-k}|^4
\exp\left(-4|\Gamma_k|nT/\hbar\right)}{|p_{+k}|^4} \right), \quad k
> k^{\ast} \nonumber\\ \label{10state}
\end{eqnarray}
Thus the steady state value of $\chi_{k}$ remains close to unity for
all $k$ near $\omega_m^{\ast}$ (Fig.\ \ref{fig3}).
Consequently $g(nT) \sim 0$. We note that the oscillations are
absent for all $n$ since the amplitude of such oscillations depends
on $|p_{+k}p_{-k}|^2$ and is vanishingly small for all $k$.

Finally for the initial state $\prod_k (1,1)^T/\sqrt{2}$, we find
$\mu_{\pm k}= (p_{\pm k}^{\ast} + q_{\pm k}^{\ast})/\sqrt{2}$. Using
this, one obtains
\begin{eqnarray}
\chi_k(nT) &=& \frac{1}{{\mathcal D}_{k}^2(\pi/4)} \Big(
\sum_{s=\pm} [ |R_{sk}|^2 e^{2 s \Gamma_k nT/\hbar}]  \nonumber\\
&&+ 2 R_{+ k} R_{-k} \cos( 2 \epsilon_k nT/\hbar) \Big).
\nonumber\\
R_{s k} &=& \frac{1}{2} [1 + \left( p_{s k}^{\ast} q_{s k} + {\rm
h.c.} \right)] \label{vk2eq}
\end{eqnarray}
We note that at large $n$, this yields
\begin{eqnarray}
\chi_k &\sim& R_{-k}\left[ 1+ \left(\frac{R_{+k} e^{-2|\Gamma_k|
nT/\hbar}}{R_{-k}} \right)^2 \right] \quad k< k^{\ast}
\nonumber\\
&\sim & R_{+k} \left[ 1+ \left(\frac{R_{- k} e^{-2 |\Gamma_k|
nT/\hbar}}{R_{+k}} \right)^2 \right], \quad k
> k^{\ast} \nonumber\\
\label{11state}
\end{eqnarray}
For $\omega_D= \omega_m^{\ast}$,  $p_{s k}^{\ast} q_{s k} \simeq 0$
for all $k$ leading $R_{\pm k} \sim 1/2$ (Fig.\ \ref{fig3}). Thus
the steady state value of $\chi_k \sim 1/2$ for all $k$; this leads
to a $g \simeq -\ln 2$ in the steady state. The other features of
$g(nT)$ can be inferred from an analysis similar to those carried
out for $\theta_{0k}= 0,\pi/2$.

The nature of the steady state of the driven non-Hermitian steady
state can be further understood by studying the steady state value
of the magnetization of the driven system. We note first that the
magnetization of the driven Ising chain is given by
\begin{eqnarray}
M(nT)  &=& - \int_0^\pi \frac{dk}{\pi} N_k(nT) = \int_0^\pi
\frac{dk}{\pi} ( 1 - 2 |v_k(nT)|^2 ) \nonumber\\ \label{magising}
\end{eqnarray}
The steady state value of the magnetization, $M^{\rm st}$,  is
obtained for $n \gg {\rm Min}[J/\Gamma_k]$. Using Eqs.\ \ref{uveq},
and starting from an initial product state $\prod_k(0,1)^{T}$, which
corresponds to $M(0)=-1$, we find that
\begin{widetext}
\begin{eqnarray}
N_k(nT) &=& \frac{-1}{{\mathcal D}_{k}^2(\pi/2)} \left( \sum_{s=\pm}
\left[|q_{s k}|^2 (|p_{s k}|^2 - |q_{s k}|^2) e^{2 s \Gamma_k
nT/\hbar}\right] + 2 {\rm Re} \left[ q_{+ k} q_{- k}^{\ast} (p_{+
k}^{\ast} p_{- k} -q_{+ k}^{\ast} q_{-k}) e^{2 i \epsilon_k
nT/\hbar}\right] \right). \label{mageq1}
\end{eqnarray}
\end{widetext}
For $|\Gamma_k| nT/\hbar \gg 1$, since $\Gamma_k$ changes sign at
$k= k^{\ast}$, the steady state value of $N_k$, $N_k^{{\rm st}}$, is
given terms of $p_{\pm k}, q_{\pm k}$ by
\begin{eqnarray}
N_k^{{\rm st}} &=& (|q_{- k}|^2 -|p_{-k}|^2) \left(1 - \eta_{0k}e^{-4 |\Gamma_k| nT/\hbar} \right) \quad k < k^{\ast}\nonumber\\
&=& (|q_{+ k}|^2 -|p_{+ k}|^2) \left(1 -  \eta_{0k}^{-1} e^{-4
|\Gamma_k|nT/\hbar} \right) \quad k > k^{\ast} \nonumber\\
\eta_{0k} &=& \frac{(|q_{+ k}|^2 -|p_{+k}|^2)}{(|p_{- k}|^2
-|q_{-k}|^2} \frac{|q_{+ k}|^2}{|q_{-k}|^2}  \label{msteady}
\end{eqnarray}
Thus when $\omega_D \sim \omega_m^{\ast}$, $N_k^{{\rm st}} \sim -1$
for all $k$ leading to a steady state magnetization $M^{\rm st} \sim
1$. However, away from these frequencies, both $p_{\pm k}$ and
$q_{\pm k}$ are finite around $k= k^{\ast}$; thus the value of
$N_k^{\rm st}$ deviates from $-1$ when $k$ is within this range.
This in turn leads to lower value of $M^{\rm st}$ when $\omega_D$ is
away from $\omega_m^{\ast}$. We therefore expect non-monotonic
behavior of $M^{\rm st}$ as a function of the drive frequency.

\begin{figure}
\rotatebox{0}{\includegraphics*[width= 0.49 \linewidth]{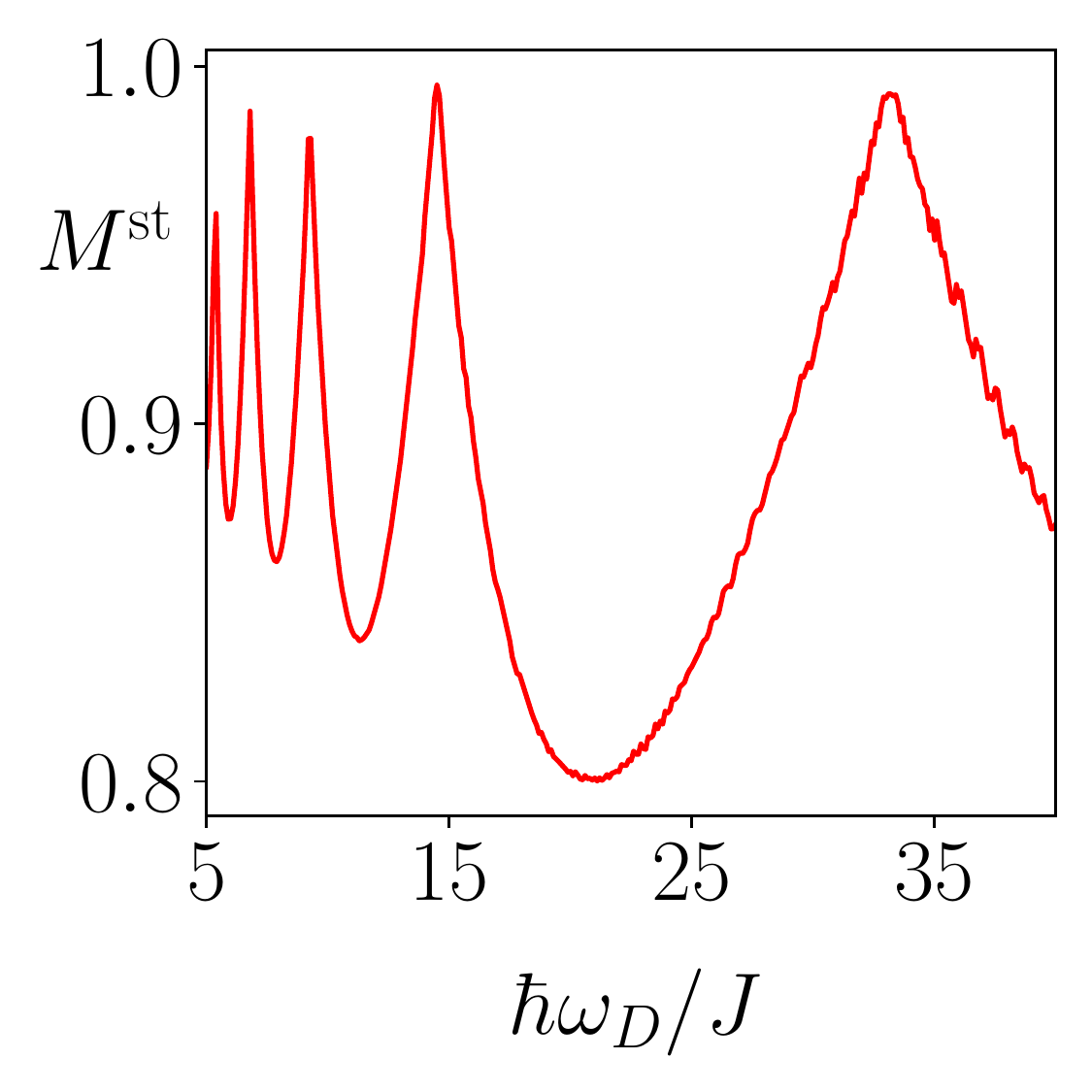}}
\rotatebox{0}{\includegraphics*[width= 0.49 \linewidth]{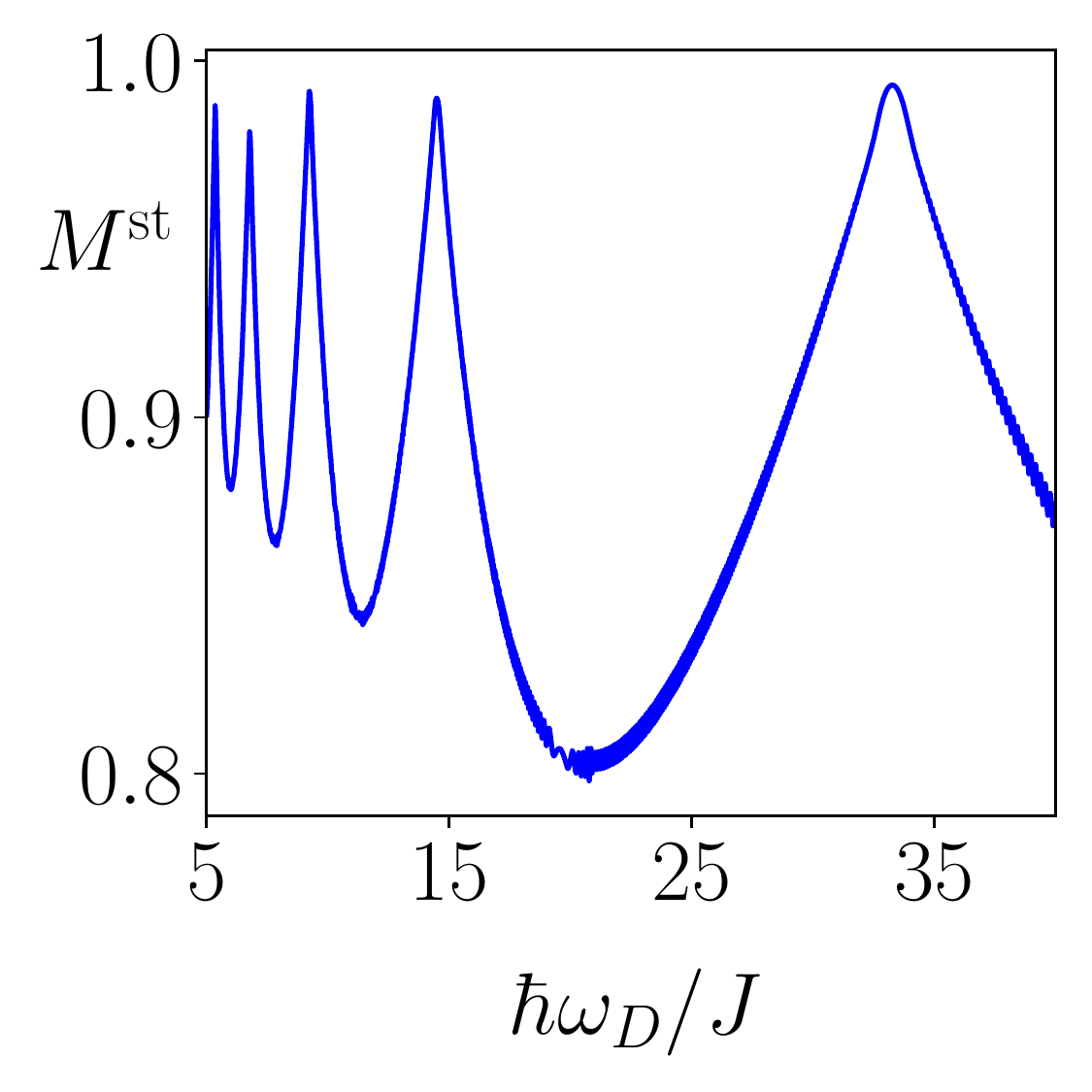}}
\caption{Top Left Panel: Plot of $M^{\rm st}$ as a function $\hbar
\omega_D/J$ as obtained from exact numerics (left panel) and second
order FPT (right panel). All parameters are same as in Fig.\
\ref{fig1}. See text for details.} \label{fig4}
\end{figure}

The plot of the steady $M^{\rm st}$, obtained from the value of
$M(nT)$ around $n\sim 1000$ after averaging over $50$ drive cycles,
plotted as a function of $\omega_D$ in the left panel Fig.\
\ref{fig4}, conforms this behavior. The plot clearly shows that the
steady state magnetization exhibits distinct dips at
$\omega_D=\omega_m^{\ast}$. The right panel of Fig.\ \ref{fig4}
shows the plot of $M^{\rm st}$ as obtained from second order
perturbation theory. Here the steady state is constructed, for each
$k$, from the normalized wavefunction $|\psi_k (nT)\rangle$ (Eq.\
\ref{normfn}) by retaining terms in $u_k(nT)$ and $v_k(nT)$ (Eq.\
\ref{uveq}) with $\Gamma_k>0$ which survive in the limit $n \to
\infty$. This yields $u_k^{\rm st}$ and $v_k^{\rm st}$ and leads to
$M^{\rm st}= \int_0^{\pi} (dk/\pi) ( 1 - 2|v_k^{\rm st}|^2 )$. The
result obtained from second order FPT in this manner is remarkably
close to the exact result. Thus we conclude that the steady state of
the driven chain bears the signature of the approximate dynamical
symmetry.

\begin{figure}
\rotatebox{0}{\includegraphics*[width= 0.49 \linewidth]{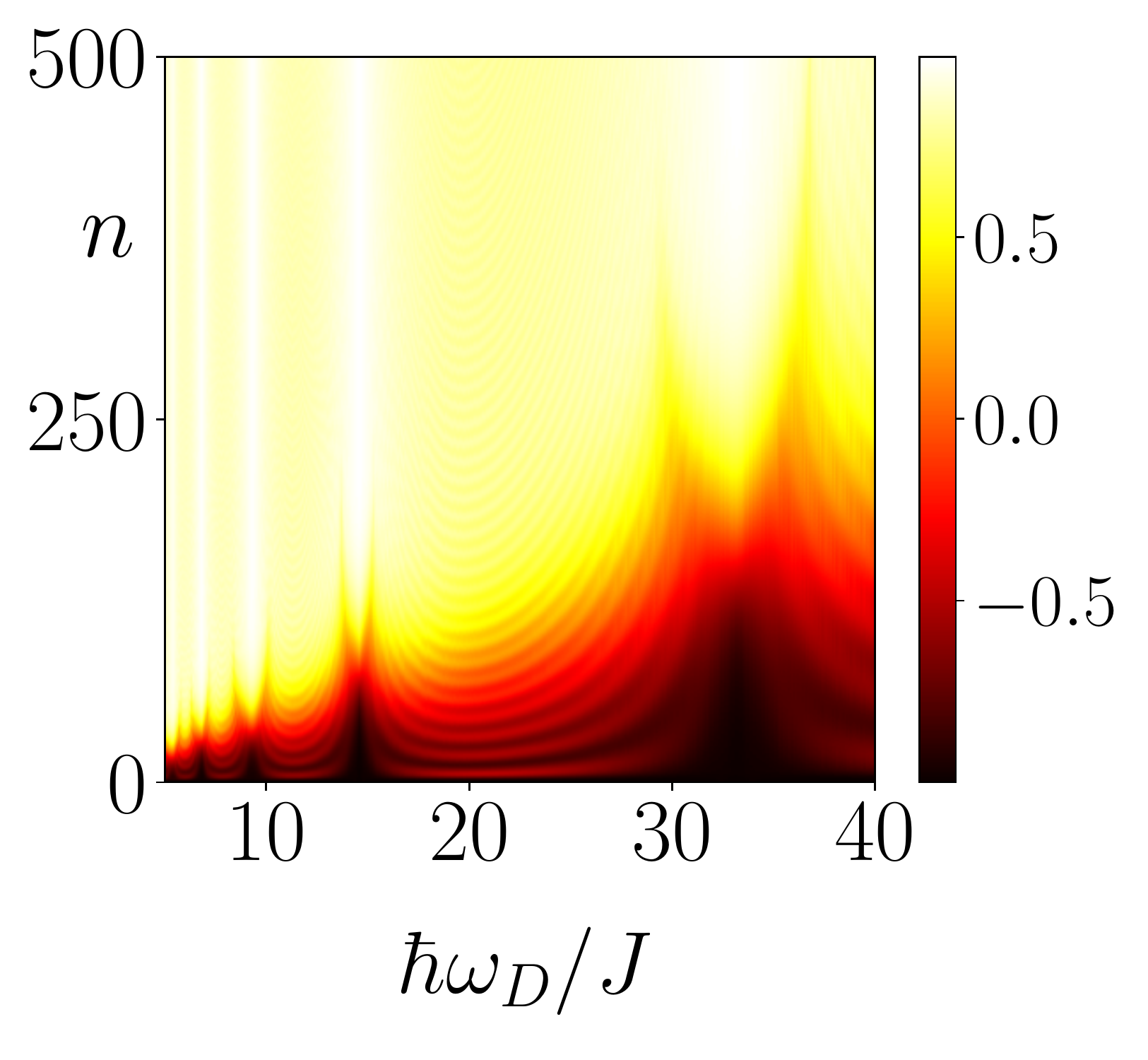}}
\rotatebox{0}{\includegraphics*[width= 0.49 \linewidth]{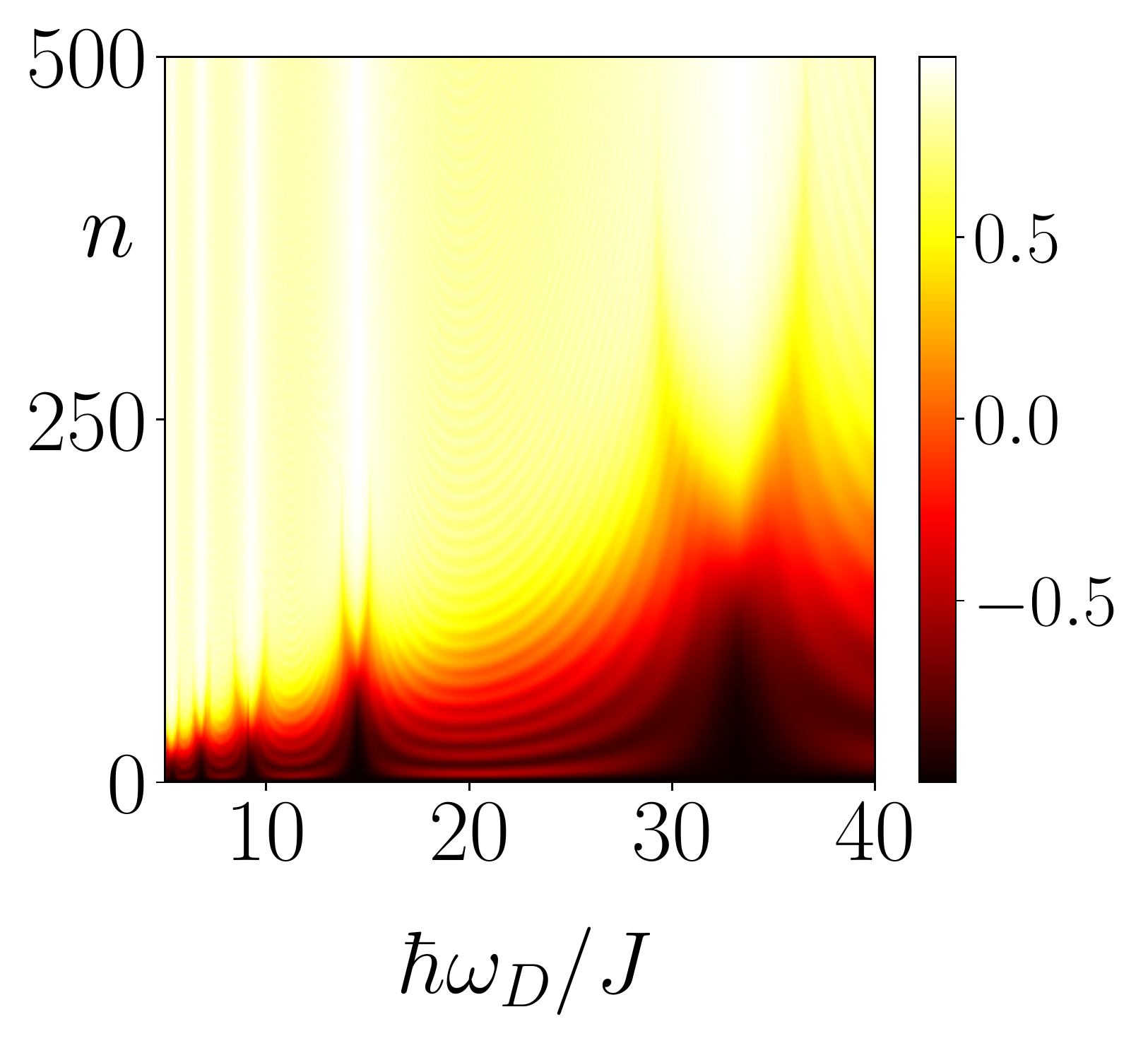}}
\caption{Top Left Panel: Plot of $M(nT)$ as a function of $n$ and
$\hbar \omega_D/J$ as obtained from exact numerics (left panel) and
second order FPT (right panel). All parameters are same as in Fig.\
\ref{fig1}. See text for details.} \label{fig5}
\end{figure}

\begin{figure}
\rotatebox{0}{\includegraphics*[width= 0.49 \linewidth]{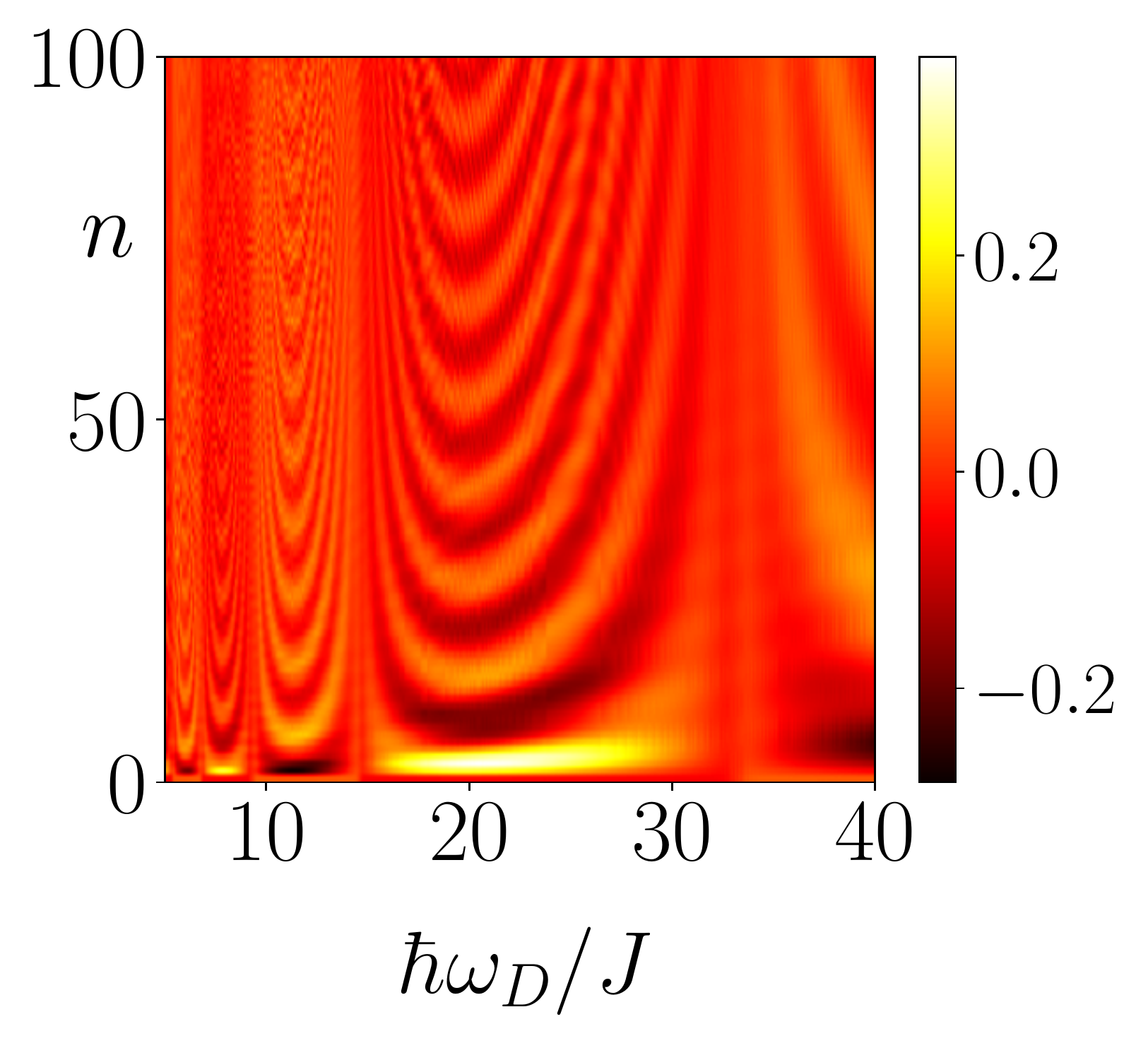}}
\rotatebox{0}{\includegraphics*[width= 0.49 \linewidth]{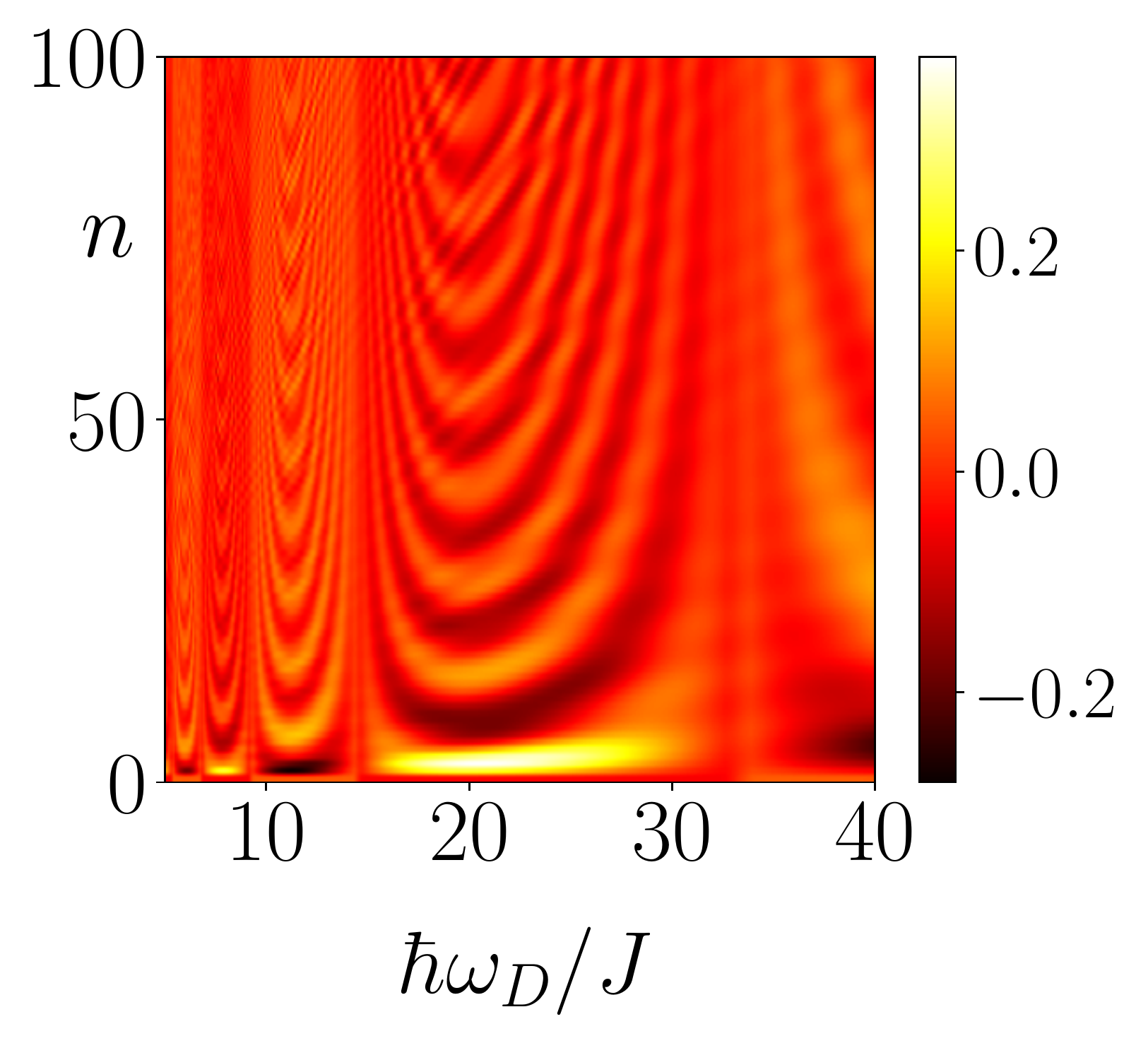}}
\caption{Top Left Panel: Plot of $F(nT)$ as a function of $n$ and
$\hbar \omega_D/J$ as obtained from exact numerics (left panel) and
second order FPT (right panel). All parameters are same as in Fig.\
\ref{fig1}. See text for details.} \label{fig6}
\end{figure}

Next, we study the behavior of the magnetization $M(nT)$ (Eq.\
\ref{magising}) as a function of $n$ and the drive frequency
$\omega_D$. The corresponding plot is shown in Fig. \ref{fig5}. The
left panel of Fig.\ \ref{fig5} shows the behavior of $M(nT)$
obtained from exact numerics while the right panel shows the
corresponding results from second order FPT; the latter sows
excellent match with the former for a wide range of drive
frequencies. The behavior of $M(nT)$ shown in these plots can be
understood from Eq.\ \ref{mageq1} as follows.

First, we note that near $\omega_m^{\ast}$, the oscillatory terms in
Eq.\ \ref{mageq1} vanishes since $p_{s k} q_{s k} \sim 0$ for
$s=\pm$ and all $k$. Thus we expect the oscillatory behavior of
$M(nT)$ to be present only away from these frequencies. This
behavior is confirmed by plots in Fig.\ \ref{fig5}. Second, from
Fig.\ \ref{fig5}, we find that at $\omega_D= \omega_m^{\ast}$, $M$
stays close to its initial value for a large number of drive cycles;
this is followed by a sharp decay to the steady state value $M^{\rm
st} \simeq 1$. The sharpness of this decay is a consequence of sharp
change of $q_{\pm k}$ and $\Gamma_k$ around $k=k^{\ast}$. Third, the
deviation of $M$ from the steady state value occurs when  $ \eta_{0k}^{-1} (\eta_{0 k}) \exp[-4 |\Gamma_k| n_c T/\hbar]
\sim 1 $ for $k >(<) k^{\ast}$ (Eq.\ \ref{msteady}). Thus the value
of $n_c$ at which this crossover occurs is exponentially sensitive
to the distribution of $|\Gamma_k|$ as a function of $k$ around $k=
k^{\ast}$. Since a sharp change of sign of $\Gamma_k$ around $k=
k^{\ast}$, which occurs around $\omega_m^{\ast}$, indicates a larger
value of $|\Gamma_k|$ for most $k$, we find that the system reaches
its steady state for smallest value of $n_c$ at $\omega_m^{\ast}$.
As one moves away from $\omega_m^{\ast}$, $n_c$ increases;
concomitantly, $q_{\pm k}$ develop finite value for larger range of
$k$ around $k^{\ast}$. Thus $M(nT)$ starts to change with $n$ for $n
<n_c$ in an oscillatory manner. The oscillation amplitude are small
near $\omega_m^{\ast}$; thus the system shows very slow change in
magnetization in the region $0 \le n \le n_c$. This leads to
peak-like structures around $\omega_m^{\ast}$ (Fig.\ \ref{fig5})
where the systems shows slow but non-zero change in the
magnetization before reaching the steady state.

Next, we study the off-diagonal fermion correlation function $F(nT)$
given by
\begin{eqnarray}
F(nT) &=& \int_0 ^{\pi} \frac{dk}{\pi} F_k(nT)  \label{odcor}
\end{eqnarray}
The plot of $F(nT)$ as a function of $n$ and $\hbar\omega_D/J$ is
shown in Fig.\ \ref{fig6}. Once again we find that the second order
FPT (right panel of Fig.\ \ref{fig6}) reproduces all the qualitative
features obtained using exact numerics (left panel of Fig.\
\ref{fig6}). To understand these features, we first note that
starting from an initial state $\prod_k (0,1)^{T}$, the expression
of $F_k(nT)$ can be written terms of $p_{\pm k}$ and $q_{\pm k}$ as
\begin{widetext}
\begin{eqnarray}
F_k(nT) &=& \frac{2}{{\mathcal D}_{k}^2(\pi/2)} \sum_{s=\pm} \left(
|q_{s k}|^2 {\rm Re}[p_{sk}^{\ast} q_{s k}] e^{2 s \Gamma_k
nT/\hbar} + |q_{-s k}|^2 \{ {\rm Re} [p_{s k}^{\ast} q_{s k}] \cos(2
\epsilon_k nT/\hbar) - s {\rm Im} [p_{s k}^{\ast}
q_{s k} ] \sin (2 \epsilon_k nT/\hbar)\} \right). \nonumber\\
\label{fkeq1}
\end{eqnarray}
\end{widetext}
From Eq.\ \ref{fkeq1}, we find that $F_k \sim 0$ for all $k$ at
$\omega_m^{\ast}$ since $|p_{sk}^{\ast}q_{sk}| \sim 0$ for $s=\pm$
and at all $k$ at these frequencies. The amplitude of the
oscillations of $F_k(nT)$ is also small for the same reason.
Consequently, $F(nT)$ remains close to zero at these frequencies for
all $n$. In contrast, significant oscillations are seem away from
$\omega_m^{\ast}$ where both $p_{sk}$ and $q_{sk}$ are finite for a
range of $k$ around $k^{\ast}$.

Thus, we find that all correlations and the fidelity bear signature
of the approximate dynamical symmetry that emerges as
$\omega_m^{\ast}$. The footprint of this emergent symmetry
constitutes lack of oscillatory features in fidelity and correlation
functions which can be discerned most easily by measuring
magnetization of the driven chain.

\subsection{Entanglement}

\begin{figure}
\rotatebox{0}{\includegraphics*[width= 0.49 \linewidth]{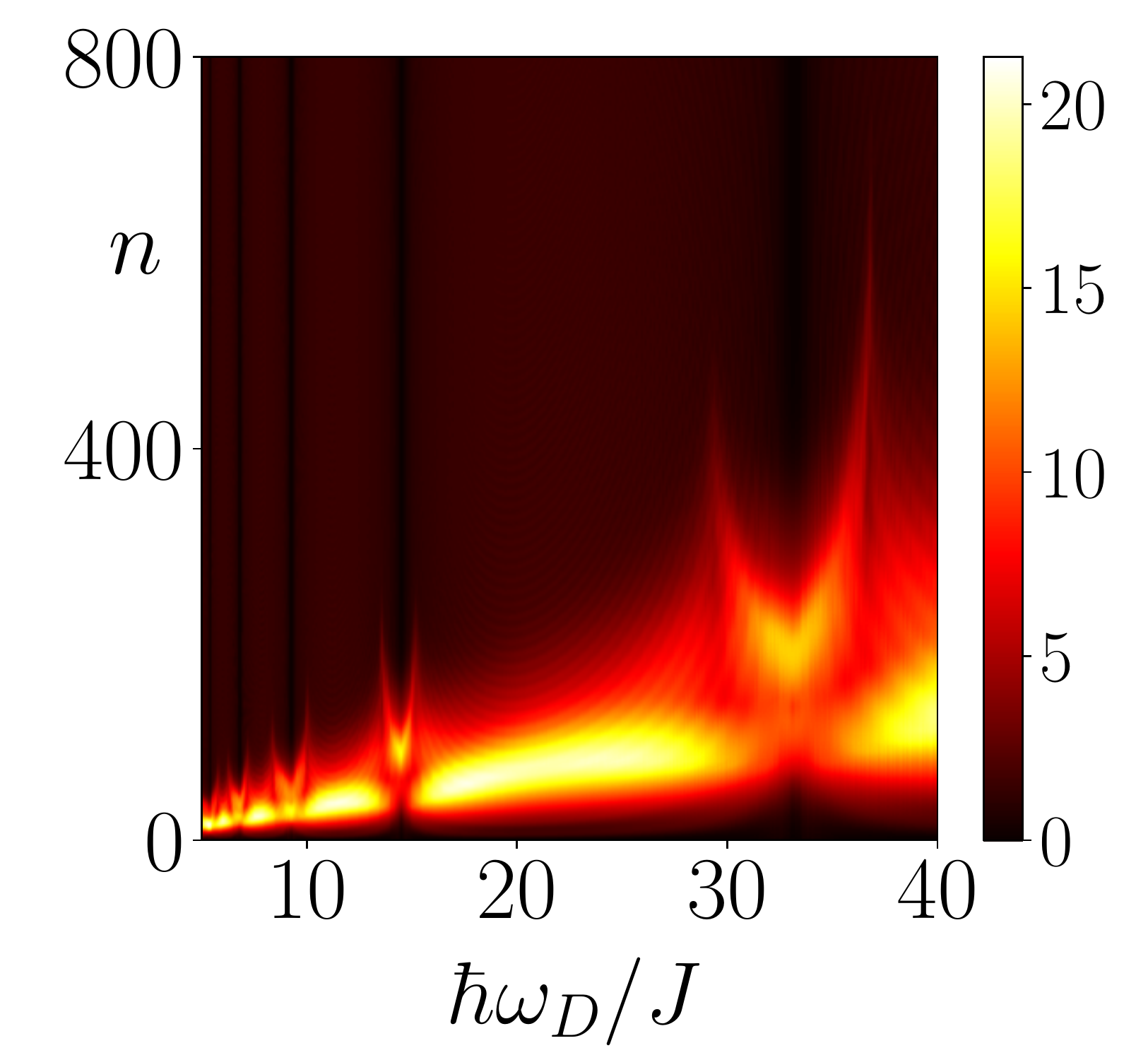}}
\rotatebox{0}{\includegraphics*[width= 0.49 \linewidth]{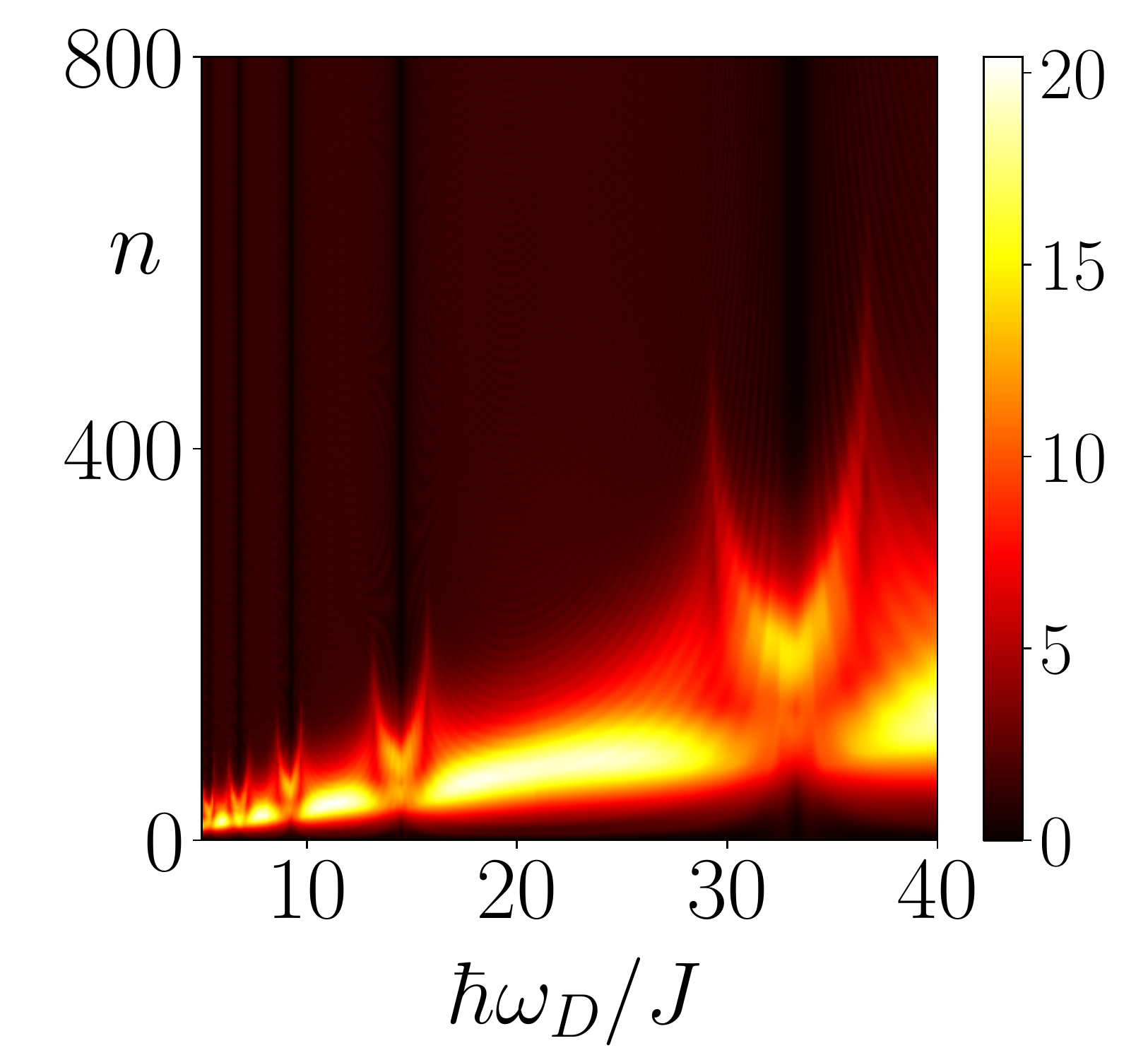}}
\rotatebox{0}{\includegraphics*[width= 0.49 \linewidth]{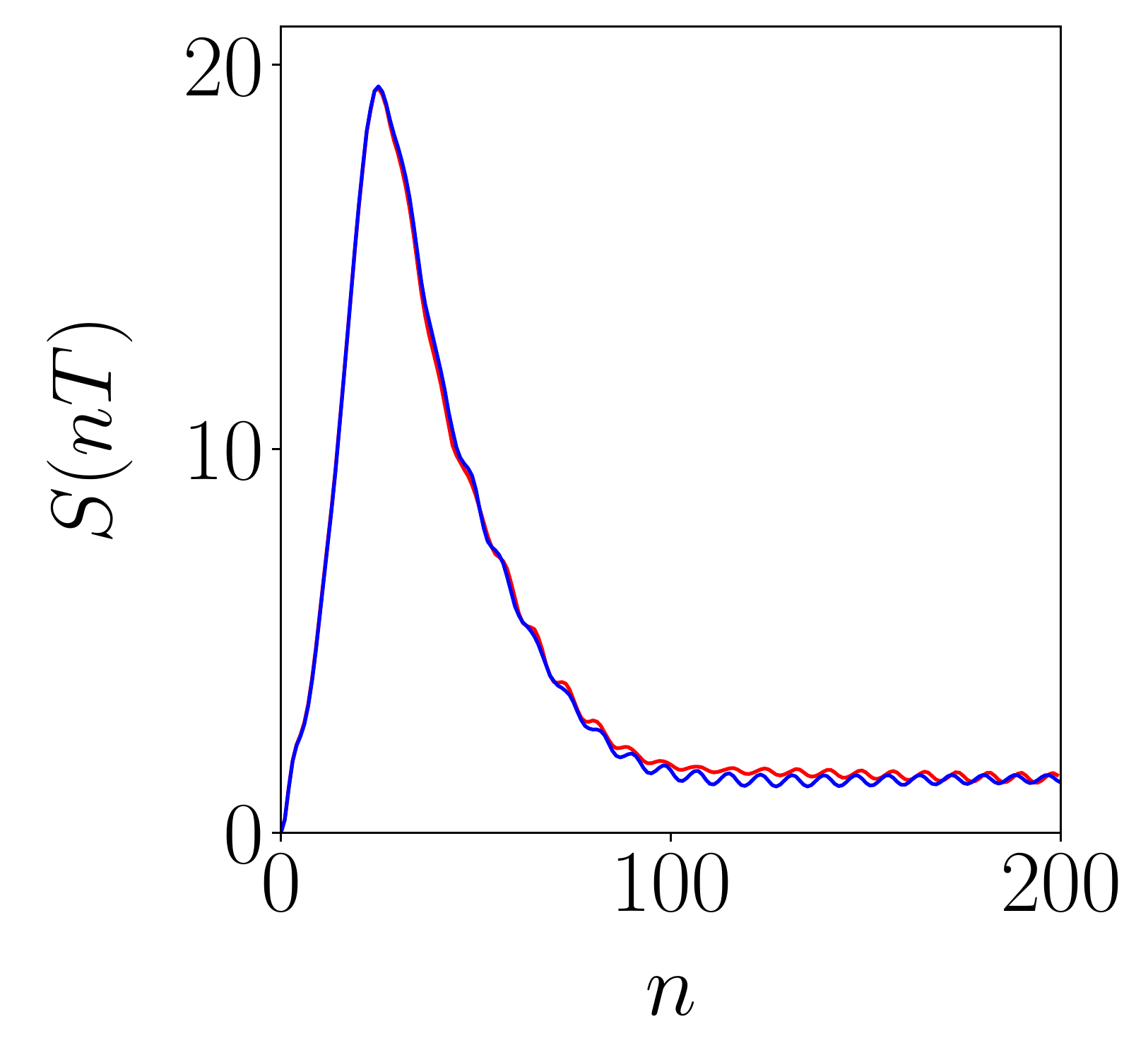}}
\rotatebox{0}{\includegraphics*[width= 0.49 \linewidth]{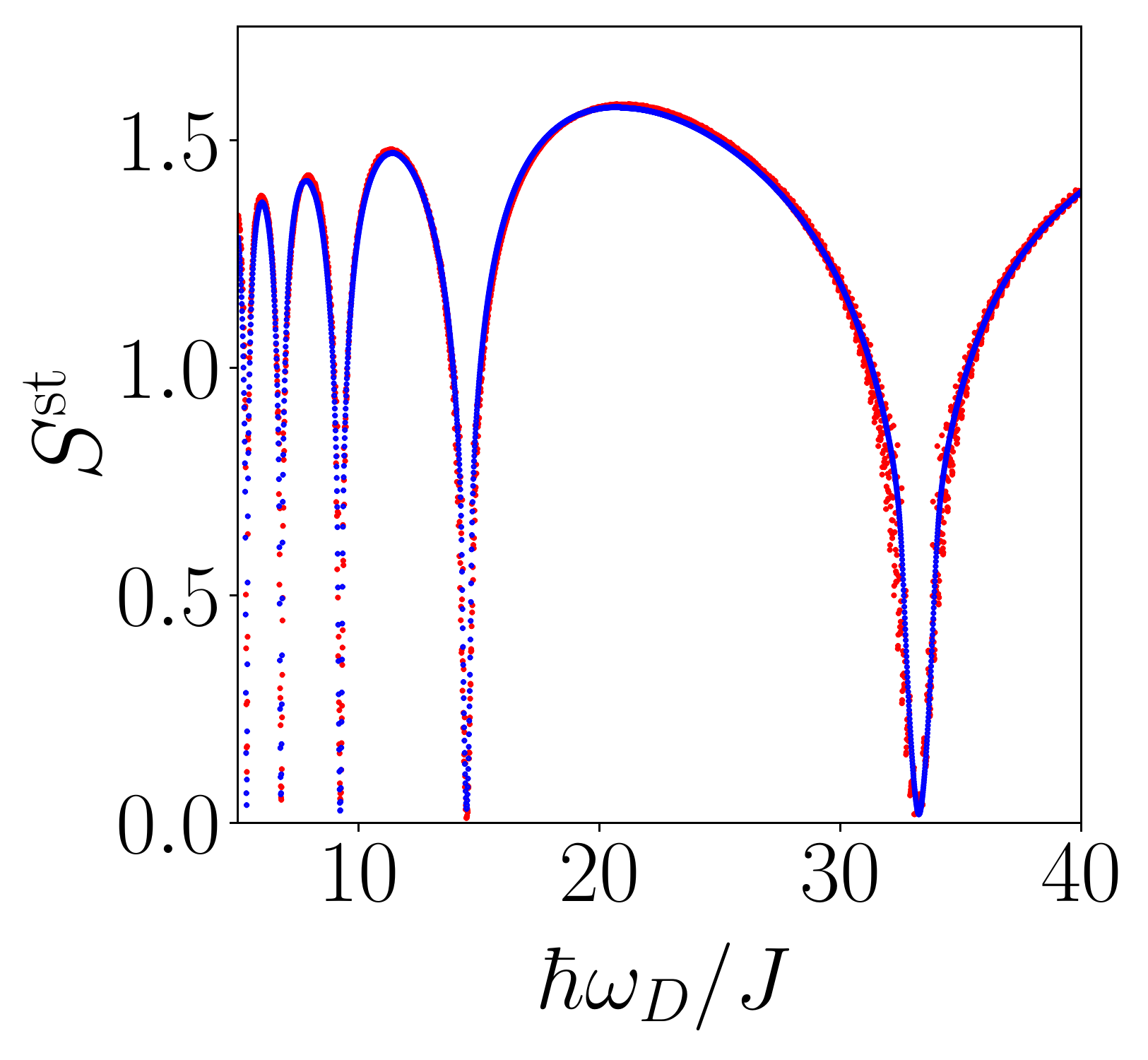}}
\caption{Top Panels: Plot of $S(nT)$ as a function of $n$ and $\hbar
\omega_D/J$ as obtained from exact numerics (top left panel) and
second order FPT (top right panel). Bottom left panel: Plot of
$S(nT)$ as a function of $n$ for $\hbar \omega_D/J=8$. Bottom right
panel: Plot of the steady state entanglement $S^{\rm st}$ as a
function of $\omega_D$ showing dips at $\omega_D=\omega_m^{\ast}$.
For the bottom panel plots red(blue) represents results obtained
from exact numerics (second order FPT). We have chosen $L=100$ for
all plots; the rest of the parameters are same as in Fig.\
\ref{fig1}. See text for details.} \label{fig7}
\end{figure}

In this section we present our results for entanglement entropy of
the driven system. In what follows, we shall mostly concentrate on
the half-chain Von-Neumann entropy $S_{\ell=L/2}(nT) \equiv S(nT)$
(Eq.\ \ref{entangexp}), where $L$ is the chain length, as a function
of $n$ and $\omega_D$.

A plot of $S(nT)$ is shown, starting from an initial state
$|\psi_0\rangle= \prod_k (0,1)^{T}$, as a function of $n$ and
$\omega_D$ in Fig.\ \ref{fig7}. The plots show that $S$ follows an
almost similar pattern as the correlation functions and hence bears
a signature of the special frequencies. Moreover, from these plots,
we find, comparing the top left and right panels of Fig.\
\ref{fig7}, that the second order FPT matches well with exact
numerics for a wide range of drive frequency.

The plot of $S$ as a function of $n$ for a fixed drive frequency
$\hbar \omega_D/J=8$ is shown in the bottom left panel of Fig.\
\ref{fig7}. The behavior of $S$, as shown in this plot, brings out a
key difference between it and its counterpart for driven Hermitian
Ising chains \cite{dtran2}. For periodically driven Hermitian
chains, $S$ is know to first increase and then saturate with
increasing $n$. In contrast, for a driven non-Hermitian chain $S$
first increases, reaches a peak, and then decays to its steady state
value at large $n$. This behavior can be understood as follows.

The initial state of the system $|\psi_0\rangle$ is a product state
leading to $S(0)=0$. For small $n$, the behavior of $S$ shows a
similar increase as in Hermitian driven chain. However, for large
$n$ where $|\Gamma_k| nT/\hbar \ge 1$ for all $k$, it start to
approach its steady state value. In contrast to driven Hermitian
chains, the steady state here has a low entropy, being an almost
product state. This indicates that $S(nT)$ for large $n$ is also
small; in fact, it approaches zero as $\omega_D \to \omega_m^{\ast}$
where the steady state is a perfect product state with $S=0$. This
ensures that $S(nT)$ is necessarily a non-monotonic function of $n$.
In between, $S(nT)$ reaches its peak value; the position of this
peak depends on both $\omega_D$ and $\gamma$. We note that these
features of $S(nT)$ are accurately captured by the second order FPT
(blue curve in the left bottom panel of Fig.\ \ref{fig7}) which
provides a near-exact match with exact numerical results (red curve
in the left bottom panel of Fig.\ \ref{fig7}).

The behavior of $S(nT)$ for $\omega_D=\omega_m^{\ast}$ as a function
of $n$ is qualitatively similar to that shown in the bottom left
panel of Fig.\ \ref{fig7} with two important differences. First, the
oscillatory features of $S$ are absent at these frequencies and
second, the steady state value of $S$ approaches zero. The latter
can be most easily inferred from the plot of $S^{\rm st}$ as a
function of $\omega_D$ as shown in the bottom right panel of Fig.\
\ref{fig7}. We find that the special frequencies $\omega_m^{\ast}$
can be distinguished by dips in $S^{\rm st}$; this can be understood
as a consequence of the fact that the steady state, at these
frequencies, are very close to the product state $|\psi_s\rangle=
\prod_k (1,0)^{T}$.  This confirms that the steady state
entanglement also bears the signature of the emergent approximate
conservation.

\section{Discussion}
\label{diss}

In this work, we have studied a class of driven 1D non-Hermitian
integrable free fermionic models in the high drive-amplitude regime.
We have identified the presence of approximately conserved
quantities that leave their imprints on the dynamics of these
models.We have shown in the appendix that such emergent conservation
can also be seen for discrete drive protocols; this demonstrates the
general, protocol-independent, nature of this phenomenon.

For a continuous drive protocol, we have used Floquet perturbation
theory to obtain the Floquet Hamiltonian $H_F$ of the driven models.
The method uses inverse of the drive amplitude as the perturbation
parameter and thus provides reasonably accurate results for high and
intermediate drive frequency regimes. This distinguishes it from the
standard high frequency expansions where the inverse frequency is
taken as the perturbation parameter. We show, using the example of
1D transverse field Ising chain, that the dynamics obtained using
$H_F$ computed from second order FPT reproduces all the features of
its exact numerical counterpart.

The Floquet Hamiltonian obtained using this method provides analytic
understanding of the reason for the emergent approximate
conservation at special frequencies. At these frequencies, whose
analytic expressions can be obtained using FPT, the first order
Floquet Hamiltonian (obtained using FPT) commutes with certain
operators. A specific example of such an operator is shown to
correspond to the transverse magnetization of the driven
non-Hermitian Ising chain. Such a conservation is approximate and it
is shown to be violated by higher (second) order Floquet
Hamiltonian. Nevertheless, this approximate emergent conservation
leaves its signature on the dynamics of the driven chain. In this
respect, non-Hermitian systems differ qualitatively from their
Hermitian counterparts studied in Refs.\ \onlinecite{ad1}; for
example, the magnetization of the latter stays very close to its
initial value for a very large number of drive cycles at such
special drive frequencies. In contrast, the magnetization of
integrable non-Hermitian systems studied here exhibit distinct
dynamics and approaches its steady state value after $n \sim 200$
drive cycles.

We discus the dynamical signature of this approximate conservation
and show that it also shapes the nature of the steady states of
these driven systems. Using the Ising model in a transverse field as
example, we show that the steady state of the driven non-Hermitian
Ising chain coincides with an eigenstate of the transverse
magnetization at these special frequencies. Moreover, the approach
of the system to the steady state shows distinct behavior at these
special frequencies; they lack the transient oscillations which is
normally present when the drive frequency is away from these special
frequencies. Such a qualitatively different behavior is also
reflected in the entanglement entropy $S$ of such systems. In
particular, for the Ising chain, the steady state entanglement
entropy $S^{\rm st}$ approaches zero at these special frequencies;
in contrast, it is finite at other drive frequencies. Our study also
indicates the non-monotonic behavior of $S$ as a function of $n$ and
ties it to the non-Hermitian nature of these models.

There have been several suggestions of realization of non-Hermitian
Ising chains \cite{daley1,dalibard1,lu1,chen1}. Some of these
protocols involve coupling a Hermitian Ising chain with a
continuously measuring devise which measures the transverse
magnetization; the effective Hamiltonian of the system in the
so-called no-click limit is then given by a Ising chain with an
imaginary component $\gamma_0$ of the transverse field
\cite{nhdyn5}. The net transverse field acting on the Ising spin
thus becomes $B+i \gamma_0$, where $B$ denotes the existing
transverse field of the uncoupled Hermitian Ising chain.  Our
proposition is to drive the chain with a time dependent magnetic
field $B=B_0+ B_1 \cos \omega_D t$ starting from an all-down spin
state. In the limit of large drive amplitude, we predict that the
steady state magnetization per unit length of the chain would be
close to $\hbar/2$ at special drive frequencies
$\omega_D=\omega_m^{\ast}$. These frequencies are predicted to be
related to the drive amplitude by $ h_1/(\hbar \omega_m^{\ast}) =
\rho_m$ where $\rho_m$ denotes the position of the $m^{\rm th}$ zero
of $J_0$ and $h_1 =\mu_0 B$ where $\mu_0$ is the magnetic moment
associated with the Ising spins. A similar phenomenon would be seen
for square pulse protocol at $ h_1/(\hbar \omega_m^{\ast}) = m \pi$.
The approach of the magnetization $M(nT)$ to its steady state value
$M^{\rm st}$ as a function of the number of drive cycles $n$ can
also be measured; we predict that the evolution of $M(nT)$ will be
consistent with Fig.\ \ref{fig5} and it will show lack of transient
oscillations for $\omega_D=\omega_m^{\ast}$.

In conclusion, we have studied the Floquet dynamics of a class of
driven non-Hermitian integrable models. We have identified special
drive frequencies in these systems which leads to emergence of
approximate conservation laws. We have identified the signature of
this phenomenon in the dynamics of the driven systems and suggested
experiments which can test our theory.

\section{Acknowledgement}

KS thanks DST, India for support through SERB project
JCB/2021/000030.

\appendix*

\section{Square pulse protocol}

In this appendix, we show the presence of approximate conservation
laws in the limit of high drive amplitude for integrable
non-Hermitian free-fermionic models for a square pulse protocol. To
this end, we consider a square pulse drove protocol
\begin{eqnarray}
g(t) &=&  g_0 \quad {\rm for} \,\,t\le T/2 \nonumber\\
&=& -g_0  \quad {\rm for} \,\, t > T/2  \label{sqprot}
\end{eqnarray}
where $T$ is the time period. Substituting Eq.\ \ref{sqprot} in Eq.\
\ref{hamint}, one finds the evolution operator of the system at
$t=T$ and for a given $\vec k$ to be
\begin{eqnarray}
U_{\vec k}^{\rm sq}(T,0) &=& e^{-i H_{\vec k}^- T/(2\hbar)} e^{-i H_{\vec k}^+
T/(2\hbar)}
\nonumber\\
H_{\vec k}^{\pm} &=& (\pm g_0 -a_{3 \vec k} + i \gamma) \tau_3 +
\Delta_{\vec k} \tau_1 \label{evolk}
\end{eqnarray}
A straightforward analysis yields
\begin{widetext}
\begin{eqnarray}
U_{\vec k}^{\rm sq} &=& \left( \begin{array} {cc} \alpha_1 & \beta_1 \\
\beta_2 & \alpha_2 \end{array} \right) \nonumber\\
\alpha_j &=& (\cos \theta_{\vec k}^+ + i (-1)^j  n_{3 \vec k}^+ \sin
\theta_{\vec k}^+)(\cos \theta_{\vec k}^- + i(-1)^j n_{3 \vec k}^-
\sin \theta_{\vec k}^-) - n_{1 \vec k}^+ n_{1 \vec k}^- \sin
\theta_{\vec k}^+\sin \theta_{\vec k}^-  \nonumber\\
\beta_j &=& -i ( n_{1 \vec k}^+ \sin \theta_{\vec k}^+ (\cos
\theta_{\vec k}^- +i (-1)^j  n_{3 \vec k}^-  \sin \theta_{\vec k}^-)
+ n_{1 \vec k}^- \sin \theta_{\vec k}^- (\cos \theta_{\vec k}^+
-i (-1)^j  n_{3 \vec k}^+  \sin \theta_{\vec k}^+)) \nonumber\\
n_{1 \vec k}^{\pm} &=& \frac{\Delta_{\vec k}}{E_{\vec k}^{\pm}}, \,
\, n_{3 \vec k}^{\pm} = \frac{\pm g_0 - a_{3 \vec k} +
i\gamma}{E_{\vec k}^{\pm}}, \quad E_{\vec k}^{\pm}= \sqrt{(\pm g_0 -
a_{3 \vec k} + i\gamma)^2+\Delta^2_{\vec k}}, \quad \theta_{\vec
k}^{\pm} = E_{\vec k}^{\pm} T/(2 \hbar)  \label{ueq}
\end{eqnarray}
\end{widetext}
Eq.\ \ref{ueq} yields exact $U_{\vec k}$ for any drive frequency and
amplitude. Now we note that for large drive amplitude $g_0 \gg
a_{3 \vec k}, \Delta_{\vec k}, \gamma$, the off-diagonal terms of
$U_{\vec k}(T,0)$ vanish for special drive frequencies
$g_0/\omega_m^{\ast} = m$ where $m \in Z$. For these frequencies,
$[\tau_{3}, U_{\vec k}(T,0)] = 0$ (to leading order in $1/g_0$) for
all $\vec k$ leading to approximate conservation which is violated
only in subleading order in $1/g_0$. This violation is hence small
in the large $g_0$ limit. This demonstrates the presence of special
frequencies for the square pulse drive protocol.

\end{document}